\documentclass[runningheads]{llncs}
\usepackage[authoryear]{natbib}
\usepackage{graphicx}
\usepackage{caption}
\usepackage{subcaption}
\usepackage{wrapfig}
\usepackage[hyphens]{url}
\usepackage{xurl}
\usepackage[colorlinks=true]{hyperref}

\usepackage{amsfonts}
\usepackage{amsmath}
\usepackage{amssymb}
\usepackage{tipa}
\usepackage{placeins}
\usepackage{multirow}
\usepackage{hyperref}

\usepackage[table,xcdraw]{xcolor}
\usepackage{tabularx}
\usepackage{mathptmx}
\usepackage{xcolor}

\usepackage{paralist}
\usepackage{array}
\usepackage{cite}
\usepackage{pgf}
\usepackage{tikz}
\usepackage{pgfplots}
\usetikzlibrary{decorations.pathmorphing,decorations.text} 
\usetikzlibrary{matrix,trees,arrows,fit,shapes,positioning,decorations,backgrounds,calc}
\pgfplotsset{compat=1.14}

\pgfdeclarelayer{lowest}
\pgfdeclarelayer{background}
\pgfdeclarelayer{foreground}
\pgfsetlayers{lowest,background,main,foreground}

\usepackage{varwidth}
\usepackage{pdfpages}
\usepackage{algorithm}
\usepackage{algorithmicx}
\usepackage{algpseudocode}
\usepackage{listings}
\usepackage{todonotes}
\usepackage[normalem]{ulem}

\newcolumntype{?}{!{\vrule width 1pt}}

\newcolumntype{L}[1]{>{\raggedright\let\newline\\\arraybackslash\hspace{0pt}}m{#1}}
\newcolumntype{C}[1]{>{\centering\let\newline\\\arraybackslash\hspace{0pt}}m{#1}}
\newcolumntype{R}[1]{>{\raggedleft\let\newline\\\arraybackslash\hspace{0pt}}m{#1}}

\usepackage{xpatch}
\xpatchcmd{\thebibliography}{\small}{\small\vspace*{-1mm}}{}{}

\usepackage{listings}

\usepackage{color}
\definecolor{gray}{rgb}{0.4,0.4,0.4}
\definecolor{darkblue}{rgb}{0.0,0.0,0.6}
\definecolor{cyan}{rgb}{0.0,0.6,0.6}
\definecolor{customblue}{rgb}{0.08,0.27,0.725}

\graphicspath{{fig/}}
\usepackage[most]{tcolorbox}
\tcbset{
  highlightstyle/.style={
    colback=gray!10, 
    colframe=gray!50, 
    boxrule=0.5pt,
    arc=2mm,
    left=1mm,
    right=1mm,
    top=1mm,
    bottom=1mm,
    boxsep=2pt
  }
}
\definecolor{myblue}{RGB}{0, 0, 128} 
\hypersetup{
  citecolor=myblue,
  linkcolor=black,
  urlcolor=magenta
}
\begin{document}

\title{Diversity and Inclusion in AI: Insights from a Survey of AI/ML Practitioners}  

\author{
Sidra Malik\inst{1}
\and
Muneera Bano\inst{1}
\and
Didar Zowghi \inst{1}
}
\institute{
   CSIRO's Data61, Australia\\
  \email{\{sidra.malik, muneera.bano, didar.zowghi\} @data61.csiro.au}\\
}

\maketitle
\begin{abstract}

Growing awareness of social biases and inequalities embedded in Artificial Intelligence (AI) systems has brought increased attention to the integration of Diversity and Inclusion (D\&I) principles throughout the AI lifecycle. Despite the rise of ethical AI guidelines, there is limited empirical evidence on how D\&I  is applied in real-world settings. This study explores how AI and Machine Learning (ML) practitioners perceive and implement D\&I principles, and identifies organisational challenges that hinder their effective adoption. Using a mixed-methods approach, we surveyed industry professionals, collecting both quantitative and qualitative data on current practices, perceived impacts, and challenges related to D\&I in AI. While most respondents recognise D\&I as essential for mitigating bias and enhancing fairness, practical implementation remains inconsistent. Our analysis revealed a disconnect between perceived benefits and current practices, with major barriers including the under-representation of marginalised groups, lack of organisational transparency, and limited awareness among early-career professionals. Despite these barriers, respondents widely agree that diverse teams contribute to ethical, trustworthy, and innovative AI systems. By underpinning the key pain points and areas requiring improvement, this study highlights the need to bridge the gap between D\&I principles and real-world AI development practices.

\end{abstract}
\keywords {Diversity, Inclusion, Artificial Intelligence, Large Language Models, Survey}

\section{Introduction}\label{sec:intro}

Artificial Intelligence (AI) systems are increasingly embedded in many aspects of daily life, influencing domains such as healthcare, education, finance, and public services. Despite AI's potential to drive innovation and efficiency, there have been growing concerns surrounding Diversity and Inclusion (D\&I) in AI systems. ``\textit{Diversity} refers to the representation of the differences in attributes of humans in a group or society. \textit{Inclusion} on the other hand is the process of proactively involving and representing the most relevant humans with diverse attributes; those who are impacted by, and have an impact on, the AI ecosystem context~\citep{zowghi2023diversity}.
 
 AI systems have demonstrated instances of biases such as in facial recognition, misidentifying individuals from certain demographic groups, particularly women and people of colour~\citep{buolamwini2018gender}. They have also led to wrongful arrests and heightened surveillance of marginalised communities~\citep{tilmes2022disability}\footnote{\href{https://www.theguardian.com/uk-news/2024/dec/08/police-unlawfully-storing-images-of-innocent-people-for-facial-recognition}{Police unlawfully storing images for facial recognition}}
\footnote{\href{https://www.cbsnews.com/news/facial-recognition-misidentification-texas/}{The case of facial recognition misidentification}}. AI-powered recruitment tools have also been found to exhibit gender discrimination due to historical biases present in training data~\citep{raghavan2020mitigating, koumoutsos2022artificial, tilmes2022disability}. For instance, Amazon’s AI recruitment tool was discontinued after it was found to systematically favour male candidates\footnote{\href{https://www.reuters.com/article/us-amazon-com-jobs-automation-insight-idUSKCN1MK08G}{Amazon's recruitment tool}}. Broader concerns persist about gender disparities in AI, with studies highlighting the under representation of women in AI research and leadership roles\footnote{\href{https://time.com/7210973/women-in-the-ai-revolution/}{Under representation of women in leadership roles}}. This is because these tools depend on language models which have been shown to reinforce stereotypes, and societal biases embedded in the training data~\citep{bender2021dangers} \footnote{\href{https://www.technologyreview.com/2023/02/15/1068354/openai-chatgpt-racial-biases/}{AI chatbots generate racially biased outputs}}. These examples highlight the risk of AI systems perpetuating and amplifying existing social inequalities when D\&I principles are not adequately integrated during the design, development, and deployment of AI systems~\citep{bano2025does, zajko2022artificial}Empirical evidence has shown that biased datasets, homogeneous development teams, and opaque governance structures can embed discrimination into AI systems~\citep{buolamwini2018gender, bender2021dangers, mehrabi2019survey}.

Ethical frameworks around D\&I in AI are widely discussed~\citep{cachat2023diversity, tilmes2022disability, fjeld2020principled}, and many studies exploring the responsibility, accountability and bias in AI agents~\citep{gudmunsen2025designing,drage2024engineers}. However, there is a notable lack of empirical studies outlining how D\&I in AI is actually being implemented in real-world AI settings. Especially, the perspectives of AI/ML practitioners, those directly involved in developing, testing, and deploying AI systems, are missing. Current efforts to promote D\&I in AI often emphasise theoretical frameworks centered around broad ethical principles, without sufficient attention to practical application~\citep{cachat2023diversity,abbasgholizadeh2024edai, jobin2019global,fjeld2020principled}.
To bridge this gap, we need to understand how organisations \textit{translate} D\&I principles into practice. While many ethical frameworks and responsible AI guidelines exist globally~\citep{eu_ethics_guidelines_ai,australian_AI_standard,aida2023,iso42001,unesco2021ethics, wef2024alignment}, their adoption may not always be practical due to organisational priorities and technical constraints. To comprehensively assess how D\&I principles are operationalised within AI systems, we designed an online survey.  The survey structure focuses on demographics and three key areas, \textbf{Current Practices}, \textbf{Perceived Impacts}, and \textbf{Challenges} (See Section ~\ref{sec:RQs}). It is important to explore  if there is a divergence between practitioners' perceptions of D\&I in AI and its actual implementation within AI systems. Understanding the challenges encountered when integrating D\&I principles into practical workflows is crucial for identifying specific stages in the AI lifecycle where these issues arise. Moreover, each of the three key areas was further examined across the five pillars of AI ecosystem: Humans, Data, Process, System, and Governance~\citep{lu2023responsible, zowghi2024ai,zowghi2023diversity}. This five pillar model was chosen as it provides a holistic framework to evaluate D\&I integration across all practical domains of AI lifecycle.

The survey was designed to capture both quantitative and qualitative data to examine how D\&I principles are integrated into AI ecosystem. Our findings reveal both progress and persistent gaps in the integration of D\&I principles within AI ecosystem. Although there is recognition of the importance of D\&I in AI, practical implementation varies across different stages of the AI lifecycle. Organisations face challenges in sourcing diverse datasets, enforcing inclusive governance policies, and aligning D\&I in AI objectives with technical workflows. Furthermore, perceptions of D\&I in AI integration differ between roles, levels of experience, gender, and organisation size, highlighting the areas where improvements are needed. The findings of this study will provide a practical understanding of the industry landscape and seek to inform researchers, organisations and policymakers on where D\&I in AI efforts stand today and how they can be improved to align AI system development with principles of fairness, equity, and societal responsibility.

The paper is organised as follows:  Section 2  provides motivation and discusses existing literature on D\&I in AI within the context of five pillars of AI ecosystem. Section 3 provides the details on the research questions for this study, Section 4 describes the
methodology, Section 5 discusses the results and Section 6
provides discussion, Section 7  highlights potential threats to validity and Section 8 discusses the conclusion and proposes future directions.

\section{Motivation and Related work}\label{sec:related}
 
 Diversity and Inclusion in AI emphasises the integration of diverse perspectives and attributes at every stage of the AI lifecycle. ``Diversity refers to the representation of the differences in attributes of humans in a group or society and Inclusion is the process of proactively involving and representing the most relevant humans with diverse attributes; those who are impacted by, and have an impact on, the AI ecosystem context."~\citep{zowghi2023diversity}. AI ecosystem consists of five pillars: humans, data, system, processes, and governance~\citep{zowghi2023diversity}. Humans are categorised into people who would use the AI system and the people who design, develop and deploy the AI system. Data widely covers: what, how, why, by whom and for whom data is collected, labelled, modelled, and stored. The process pillar defines all the activities and tasks that are carried out to deliver an AI system. The process pillar is further divided into pre-development, development and post development phases. An AI system 
 typically uses large historical datasets to make predictions, recommendations or decisions. AI governance refers to the frameworks, policies, legal and regulatory requirements that oversee the development, deployment, and management of AI systems.
D\&I serves as the core foundation for building AI systems that are fair, ethical, and responsible\citep{bano2024vision}. By integrating diverse perspectives, experiences, and data across the AI-lifecycle, it helps to address biases that can lead to unfair outcomes, undermining the fairness and inclusivity of AI systems~\citep{mehrabi2019survey}. Research has shown that biases can emerge from multiple sources, including biased data, algorithmic design choices, and human decision-making during system development and evaluation~\citep{ferrara2023fairness, mehrabi2019survey}. Addressing these biases requires a comprehensive understanding of the challenges associated with D\&I in AI. In this regard, Shams et al.~\citep{shams2023ai} conducted a systematic literature review, identifying 55 unique challenges and 33 solutions related to D\&I in AI, highlighting the necessity of integrating diverse perspectives to mitigate biases and enhancing system transparency. To augment the previously conducted reviews, in this section, we present our synthesis of an open-ended grey literature search, industry surveys, and interview studies that have focused on D\&I (and Diversity, Equity and Inclusion DEI) in AI.

Upon examining the state-of-the-art, we find that D\&I in AI is often intertwined with discussions on bias and fairness with limited industry surveys or interviews on this topic. For example, in ~\citep{rakova2021where}, the authors conducted semi-structured interviews  with industry professionals to examine the organisational dynamics influencing responsible AI initiatives. The study identifies key enablers and barriers, such as organisational culture, structural support, and resource allocation, that impact the successful implementation of responsible AI practices. 

To the best of our knowledge, none of the existing studies have examined D\&I practices  in AI through the lens of the five pillars of the AI ecosystem. Therefore, we have structured our review of the literature according to these five key pillars~\citep{zowghi2023diversity}: humans, data, system,  process, and governance. Thus, the purpose of this survey is to gather insights from AI industry practitioners—including developers, managers, and policymakers—who play a pivotal role in shaping organisational practices in AI ecosystems. The insights we seek are the perceived impact of these efforts, how D\&I is currently integrated into AI development lifecycle, and the challenges organisations face - providing empirical insights to bridge the gap between theory and practice.

\subsection{Humans: Recruitment and Team Structures}
Many studies~\citep{bano2024diversity} have emphasised the need for intentional integration of D\&I principles in both the design of recruitment tools and the inclusive composition of development teams to ensure equitable and ethical outcomes.
The adoption of AI-powered tools in HR and recruitment processes enables the automation and streamlining of hiring tasks. However, these technologies also introduce unique risks, such as perpetuating existing biases, fostering digital exclusion, and enabling discriminatory practices in job advertising and candidate targeting~\citep{uk2023responsibleai}. For example, Amazon’s AI recruitment tool, which was discontinued after it was found to systematically discriminate against women in male-dominated roles~\citep{dastin2018amazon}. Another well known case for gender bias was found ~\citep{west2019discriminating} which reported women comprised only 15\% of AI research staff at Facebook and 10\% at Google. Such cases highlight the unintended consequences of algorithmic decision-making and the need for proactive bias mitigation strategies. Another recent study argues~\citep{tilmes2022disability} that addressing bias in AI requires going beyond technical fixes and instead focusing on justice, accessibility, and the lived experiences of disabled individuals.
Some steps have been taken, such as implementing tools like ``gender decoders" to analyse job postings to eliminate potentially biased language~\citep{ferrara2023fairness}. UK Government's Responsible AI in Recruitment Guide~\citep{uk2023responsibleai} highlights how AI assurance mechanisms equip organisations with tools, processes, and metrics to assess AI system performance, mitigate risks, and ensure adherence to regulatory requirements. In another study~\citep{fritts2021ai}, the authors argue that replacing human recruiters with algorithms may undermine the relational aspects of hiring, as algorithms lack the capacity for empathy and judgement. 
Beyond recruitment tools, diverse and inclusive teams have been shown to contribute significantly to the fairness and effectiveness of AI systems. Studies suggest that diverse teams are better equipped to identify and address biases, particularly those that affect underrepresented groups~\citep{west2019discriminating}. Failing to integrate diverse attributes and perspectives into AI-driven recruitment systems perpetuates unfair hiring practices, limits workforce diversity, and undermines ethical considerations~\citep{bano2024diversity}. 

Ensuring fairness in AI systems requires addressing biases that arise from multiple sources, as stated above. For example, biases stemming from biased training data or protected group disparities can impact the fairness of AI systems, making it crucial to adopt pre-processing techniques that balance inclusion with model accuracy~\citep{jui2024fairness}. Similarly, in-processing approaches help modify algorithms to account for fairness constraints during model training~\citep{jui2024fairness,mehrabi2019survey}.

\subsection{Data and Bias Mitigation}
The role of diverse datasets in ensuring fair AI outcomes cannot be overstated. Biases in data and algorithms, often rooted in societal disparities, affect fairness and equity in AI~\citep{kuhlman2020no}. Several researchers have emphasised that biased datasets are a primary source of algorithmic bias~\citep{mehrabi2019survey,obermeyer2019dissecting}. For instance, predictive models in healthcare have been found to provide less accurate recommendations for non-white patients due to under representation in the training data~\citep{obermeyer2019dissecting}. Similarly, another study on commercial facial recognition systems~\citep{buolamwini2018gender} exhibits significant bias in identifying individuals with darker skin tones, highlighting the consequences of using non-diverse datasets. Mehrabi et al.~\citep{mehrabi2019survey} conducted a literature review on the sources of data and algorithmic biases in AI applications and presented a taxonomy for fairness definitions used to avoid bias in AI. Another study~\citep{le2022survey} highlights gaps in existing datasets, including limited diversity and under-representation of certain groups, and their effect on fair and equitable ML outcomes. The study of Chen et al.~\citep{chen2024diversity} emphasises the importance of incorporating a wide range of linguistic and demographic perspectives in synthetic data to enhance the fairness of LLMs.  
Ferrara's survey~\citep{ferrara2023fairness} identifies bias sources in AI, highlights risks of generative AI reinforcing stereotypes, and emphasises interdisciplinary collaboration for fairer systems.
Ryan et.al~\citep{ryan2023integrating} conducted interviews with 18 experts from the fields of HCI and ML to explore the integration of fairness into the software design process. The findings of the study reveal that perceptions of fairness are subjective and multifaceted, varying significantly across different ML applications. They identified the key challenges, such as the complexity of understanding fairness, lack of high-quality data, and contextual factors in which ML systems are deployed. The study emphasises the necessity for collaboration between HCI and ML communities to address these challenges effectively.
Another interview study on fairness toolkits~\citep{deng2022exploring} was conducted in 2022. Fairness toolkits are collections of resources, including metrics and algorithms, designed to help developers assess and mitigate biases in ML models.  This study reveals that while these toolkits are designed to mitigate biases, practitioners often encounter difficulties integrating them into existing workflows. 

\subsection{AI lifecycle: Process and Systems}
AI systems often face inherent trade-offs between fairness, accuracy, and efficiency. For example, while fairness constraints can improve equitable outcomes, they may sometimes reduce model performance or increase computational complexity. Additionally, there is resistance to adopting D\&I practices in AI workflows due to concerns over perceived inefficiencies or misalignment with organisational priorities~\citep{bano2024diversity}. Another major challenge is the lack of transparency and interpretability in AI systems, often referred to as the ``black box" problem. Many AI models, particularly deep learning systems, operate in ways that are opaque to stakeholders, complicating efforts to identify and address biases~\citep{doshi2017towards}. The report published by the World Economic Forum, emphasises the necessity of embedding D\&I throughout the AI lifecycle~\citep{wef2022blueprint}. It provides a comprehensive framework to assist organisations in integrating D\&I principles at each stage of AI development, from data collection to deployment.  Similarly, Rabonato et al.~\citep{rabonato2024systematic} explored fairness techniques and metrics that are critical for evaluating bias at various stages of AI development.
Furthermore, organisations must invest in training programs, periodic audits, and interdisciplinary collaboration to effectively embed D\&I principles throughout the AI development and deployment lifecycle~\citep{shams2023ai}.

\subsection{Governance } 
Another challenge is the lack of institutional support for D\&I initiatives. According to a report by the AI Now Institute~\citep{west2019discriminating}, while many organisations have acknowledged the need for D\&I, few have implemented formal policies or frameworks to ensure its consistent application. To tackle the challenges of D\&I, several governance frameworks and ethical guidelines have been proposed. The European Union’s Guidelines on Trustworthy AI~\citep{eu2019trustworthy} emphasise the need for inclusive and transparent AI systems, encouraging developers to embed fairness and accountability throughout the AI lifecycle. Similarly, the US AI Bill of Rights~\citep{whitehouse2022aibill} outlines principles for ensuring that AI technologies are designed and used in ways that respect human dignity and prevent discrimination.

Corporate governance strategies are also evolving, with companies like Google and Microsoft implementing internal policies to promote D\&I in AI. Google’s Diversity, Equity, and Inclusion (DEI) Framework focuses on hiring diverse talent, implementing bias audits, and ensuring transparency in AI development~\citep{google2021diversity}. Similarly, Microsoft’s Diversity and Inclusion Strategy outlines commitments to inclusive AI by promoting underrepresented groups in AI research and development~\citep{microsoft2023diversity}.

However, a significant gap persists between policy and practice. According to Workday Global Survey on AI Trust Gap~\citep{workday2024ai}, 4 in 5 employees say their company has yet to share guidelines on responsible AI use. This disconnect highlights the need for more effective communication and implementation of governance policies to ensure that D\&I in AI principles are not only established but also actively practised within organisations.

While existing literature provides valuable insights into D\&I in AI challenges, bias mitigation strategies, and governance frameworks in AI, much of the research remains theoretical or focused on narrow aspects of AI ethics. There is limited empirical evidence on how organisations practically implement D\&I principles across the AI lifecycle and how these efforts vary across different industry sectors and practitioners' demographics. Additionally, most of the studies do not comprehensively address D\&I in AI through a structured framework encompassing the five pillars of AI ecosystem. This study seeks to bridge this gap by providing an empirical, practitioner-focused analysis of D\&I in AI ecosystem.

\section{Research Questions}
\label{sec:RQs}
In this section, we present the research questions that guide our investigation into the integration of D\&I principles within AI development practices. These questions are designed to explore practitioners' perspectives, current organisational strategies, and the challenges encountered in embedding D\&I across the AI lifecycle. As mentioned in section ~\ref{sec:intro}, the survey was structured around five key pillars to capture the broader organisational and technical factors influencing D\&I. This structure was chosen to ensure a holistic analysis, capturing both individual and organisational efforts to embed D\&I in AI development. Each research  question entails five survey questions mapping to the five pillars of AI ecosystem. For each research question, we provide the reason why exploring practitioners' answers is important. 
 \begin{itemize}

 \item  \textbf{RQ1-Perceived Impact: How do practitioners perceive the influence of Diversity and Inclusion in AI?}

Perceptions of D\&I in AI systems play a crucial role in shaping organisational priorities. While some practitioners may recognise D\&I in AI as essential for improving fairness, mitigating bias, others may view it as an added complexity with uncertain benefits. These perspectives can again differ based on industry, organisational culture, and individual roles. By examining these perceptions, this study aims to uncover how D\&I efforts impact AI development, where they provide the most value, and what factors shape practitioners' views on their importance. 

     \item \textbf{RQ2 - Current Practices: What strategies and practices are currently implemented by organisations in the AI/ML industry to foster Diversity and Inclusion in AI?}

 Understanding the current strategies and practices adopted by organisations is essential to identify areas requiring improvement. While many organisations commit to ethical frameworks, the extent to which these principles translate into practical measures varies significantly. Some organisations may prioritise inclusive hiring initiatives, while others may focus on ensuring diverse representation in data collection or establishing policies for fairness and accountability. Some organisations also value in-house training and restructuring of their teams to promote D\&I in all the areas of AI development lifecycle. However, without clear benchmarks or evaluation mechanisms, the effectiveness of these efforts remains uncertain. Additionally, the interplay between organisational culture, regulatory pressures, and business priorities further influences how D\&I initiatives are shaped and sustained in any organisation.
    
\item \textbf{RQ3 - Challenges: What are the primary challenges faced by AI/ML organisations in establishing and sustaining Diversity and Inclusion in AI?}

Identifying the key challenges organisations face in establishing and sustaining D\&I practices in AI-lifecycle is crucial. While many organisations recognise the importance of D\&I in AI, some issues like unconscious biases, resistance to change, lack of clear policies, and insufficient resources are major barriers for many organisations. Challenges may also arise from difficulties in integrating D\&I into existing workflows, measuring its impact, or aligning these efforts with broader organisational goals.

\end{itemize}

\section{Methodology}
\label{sec:method}
To conduct this survey, we sought approval from the Human Research Ethics Committee  of our organisation. To ensure compliance with ethical guidelines and data privacy regulations, the survey was administered using Microsoft Forms, a platform maintained by our organisation's  IT department. This ensured secure data collection and storage, maintaining participant confidentiality and anonymity throughout the process.
\subsection{Survey Design, Structure and Development}
This study employed a mixed-methods approach, combining quantitative and qualitative data. The survey was designed to balance breadth and depth, with 25 questions addressing demographics (see Appendix A), current D\&I practices, challenges, and perceived benefits. Survey questions were structured as multiple-choice, and open-ended responses, allowing both quantitative and qualitative insights. The survey instrument was developed through iterative consultations with experts within our research team.
\subsection{Pilot Study}
A pilot study was conducted internally among AI/ML professionals to assess the survey’s clarity, effectiveness, and completeness. Four AI/ML practitioners were included in the pilot study to give feedback on the survey design. The primary objective was to identify any ambiguities or misinterpretations in the questions, ensuring they were clearly understood by respondents. Additionally, the pilot study evaluated the average time required to complete the survey and gathered feedback on whether the provided multiple-choice options were comprehensive or if additional response categories were necessary. Insights from this pilot study allowed for refinements in question wording, response options, and overall structure, ensuring that the final survey was clear, inclusive, and well-aligned with the study’s objectives.
\subsection{Participant Recruitment and Demographics}
The survey targeted AI/ML practitioners, developers, managers, and governance professionals across various industries and organisational structures. Participants were invited through industry networks, professional associations, internal mailing lists, LinkedIn, and other social media platforms to ensure diverse representation. In total, 61 respondents participated in the survey, offering perspectives from varied demographic backgrounds, organisational sizes, and levels of experience. The survey was conducted over three months [October 2024] and [December 2024]. Participants completed the survey anonymously using MS Forms.
\subsection{Data Analysis Approach}
A mixed-methods approach was used to analyse the survey responses, combining statistical analysis for quantitative data with thematic analysis for qualitative responses. Quantitative data was examined through descriptive statistics, including frequency distributions and cross-tabulations, to identify patterns in responses across different demographics such as gender, organisational sizes, roles, experience levels and AI domains. Cross-analysis conducted for comparisons, between demographics and the significant trends in each of the sections: current practices, perceived impacts and challenges section. For example, we observed if organisation size impacted the current practices, how early-career practitioners’ perspectives on D\&I compared to those with extensive experience, differences in D\&I challenges across industries etc. 
For qualitative responses, a thematic coding process was applied, identifying recurring themes and concerns related to D\&I in AI development. Responses were categorised into key themes.

\section{Survey Results}\label{sec:results}

\subsection{Demographic Analysis}

\begin{figure}[hbt!]
\centering
\includegraphics[width=0.95\textwidth]{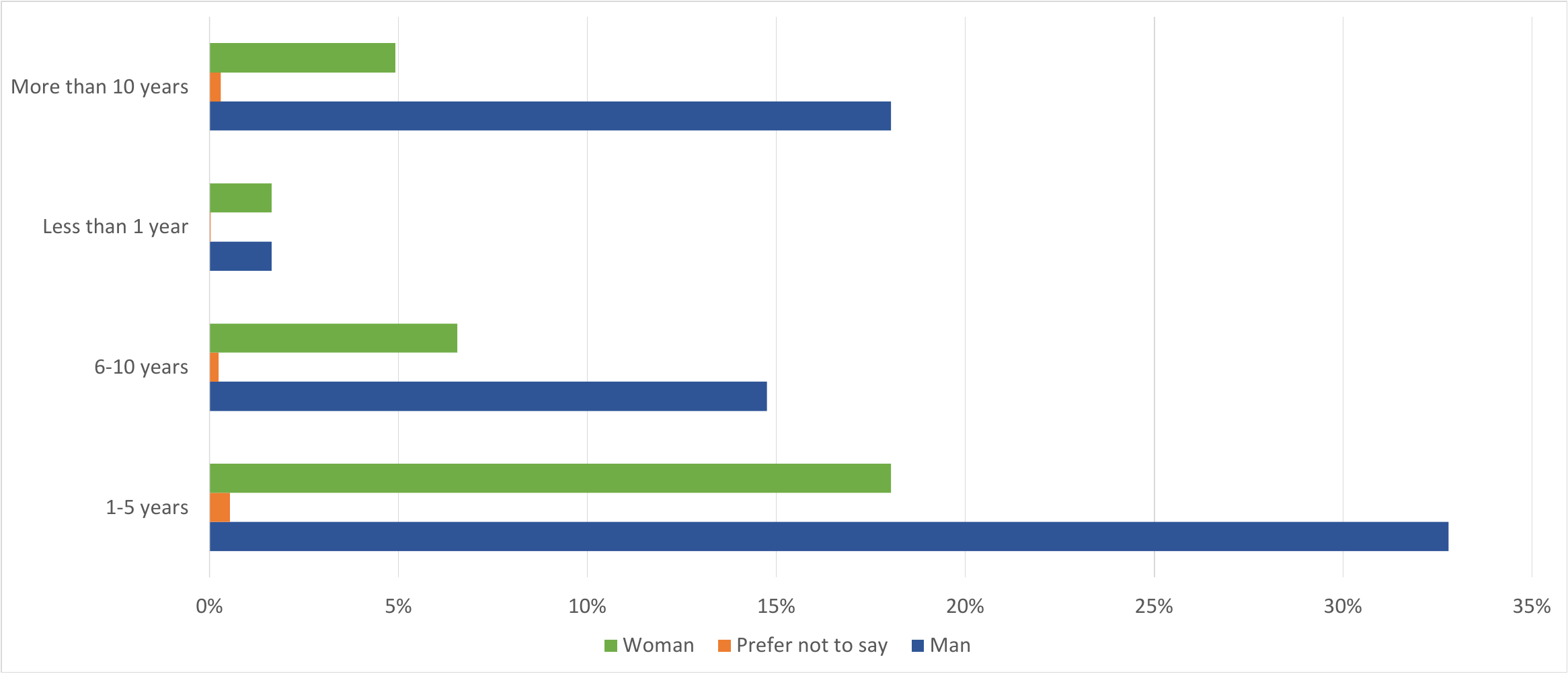}
\caption{Gender vs. Experience}
\label{Fig:gendervsexp}
\end{figure}

The survey revealed diverse demographics and work environments among respondents. 
\begin{itemize}
    \item \textbf{Gender:} 67\% identified as men, 31\% as women  and a small portion who preferred not to disclose their gender (2\%).
    \item \textbf{ Organisation Size:} Organisations varied in size, with 39\% from large organisations (more than 1000 employees) and 38\% from small companies (1-50 employees), while the rest were spread across mid-sized firms. 
    \item \textbf{Ethnicity:} 70\% of respondents identified as Asian, followed by Caucasian (13\%), Middle Eastern or North African (10\%), and minimal representations from Indigenous, Black or African Descent, and Southern European backgrounds.
    \item  \textbf{Experience: }Over half (52\%) were early-career professionals (1-5 years), with 21\% having more than 10 years of experience. 
    \item \textbf{Roles: }Most respondents worked in development roles (77\%), with notable representation in management (36\%) and governance (16\%). 
    \item \textbf{Domains: }Popular domains included Science and Research (49\%), Banking and Finance (41\%), and Health and Medical (34\%). 
    \item \textbf{AI Systems: }Large Language Models (80\%) and Predictive Models/Analytics (72\%) were the most common, followed by Classification Systems (64\%) and Image Processing/Computer Vision (51\%).
    \end{itemize}
These figures suggest that the respondents belong to a workforce heavily engaged in technical development, with a strong concentration in large-scale AI systems, yet with notable diversity gaps in ethnicity and gender. In terms of gender representation across experience levels (see Figure~\ref{Fig:gendervsexp}), men dominate the ``more than 10 years of experience" category, while most respondents, regardless of gender, fall within the 1-5 year experience range. 

Further, the cross analysis between the gender and roles suggests that gender disparities are evident in role distribution. Men overwhelmingly occupy development (74\%), management (70\%), and governance (70\%) positions. While women are better represented in development roles (30\%) compared to management (26\%) and governance (30\%), suggesting fewer women progressing into leadership positions. Analysis across ethnic groups  also reveals a clear gender imbalance, with men being the majority in most categories, particularly among Asians (64\%). Some ethnic groups, such as Middle Eastern or North African, Southern European, and Indigenous communities, have almost no female representation. Caucasian and Black professionals also have relatively low representation in the respondents' group.

\begin{figure}[hbt!]
\centering
\includegraphics[width=0.8\textwidth]{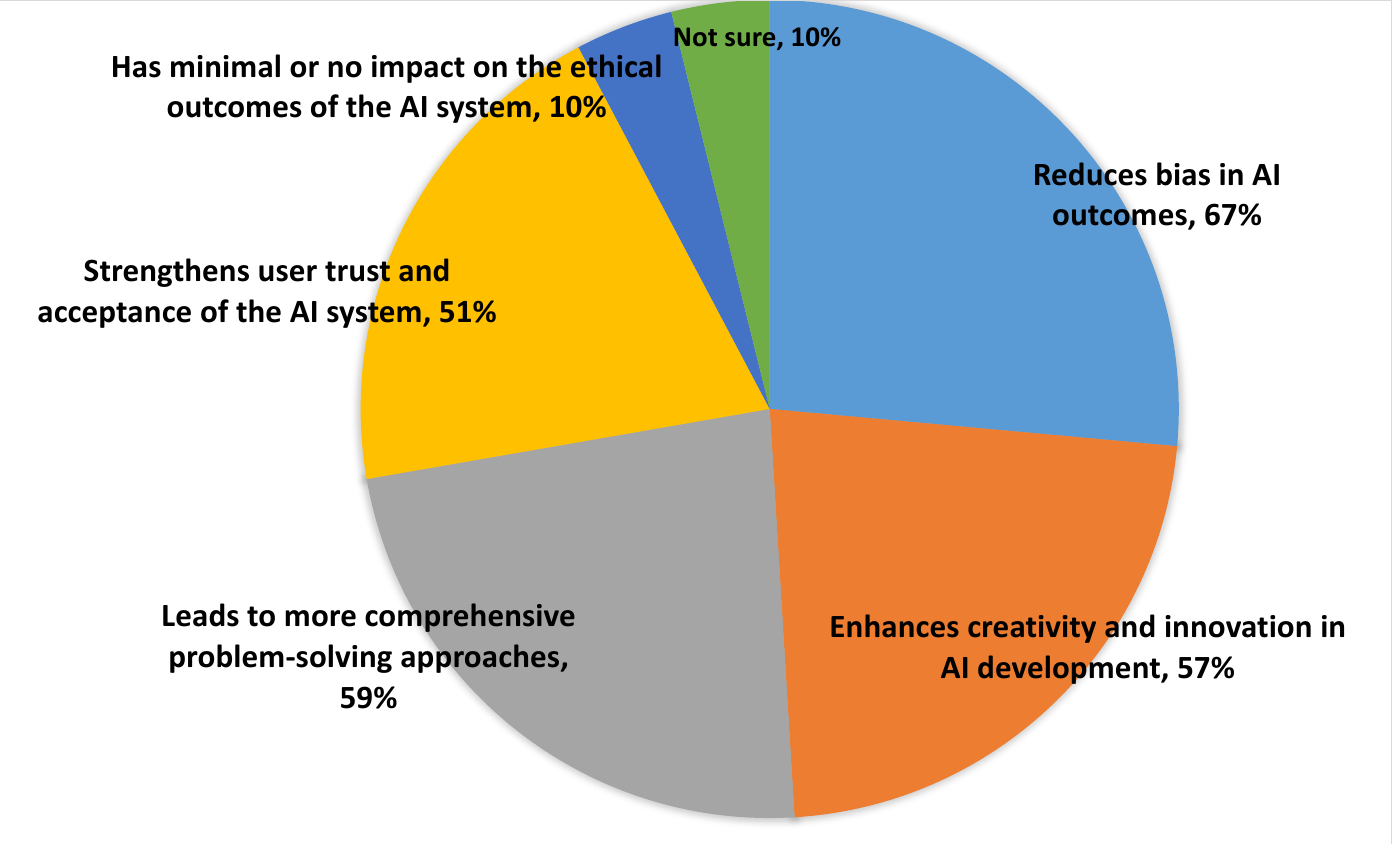}
\caption{Perceived Impact of Diverse and Inclusive Teams om D\&I in AI}
\label{Fig:PIH}
\end{figure}

\vspace{1cm}
\subsection{Perceived Impact of Diversity and Inclusion in AI}
The perception of D\&I's impact on AI varies among professionals. We first analysed the perception regarding the diverse and inclusive team at a workplace. As shown in the Figure~\ref{Fig:PIH}, a majority (67\%) believes that diverse and inclusive teams help reduce bias in AI systems, while 57\% associate it with greater creativity and innovation, and 59\% believe it to enhance problem-solving. Additionally, 51\% of respondents agree that D\&I in AI can increase user trust and acceptance. Despite these positive perceptions, a small but notable group (10\%) remains sceptical, believing that D\&I have minimal or no real impact.

The cross analysis of these 10\% sceptics' with their demographics reveal that this thought (D\&I has minimal or no impact) is common among men (67\%), particularly those from Asian (50\%) and Caucasian (33\%) backgrounds, with many falling within the 35–44 age group. Many of these respondents work in development roles, and half are employed in small organisations. A similar pattern is found among respondents who are unsure about the impact of diverse and inclusive teams, with men forming the majority (83\%). 

When considering diverse and inclusive data, most respondents (87\%) agree that it reduces AI bias, 70\% believe it helps AI generalise better across populations, and 59\% say it improves accuracy. However, 13\% believe it has little to no impact, while 5\% remain unsure. A cross analysis between the perceived impacts and demographics shows that 70\% of those who perceive minimal or no impact hold development roles. Additionally, 30\% of these sceptics come from regulated industries such as banking, finance, and healthcare.

\begin{figure}[hbt!]
\centering
\includegraphics[width=0.8\textwidth]{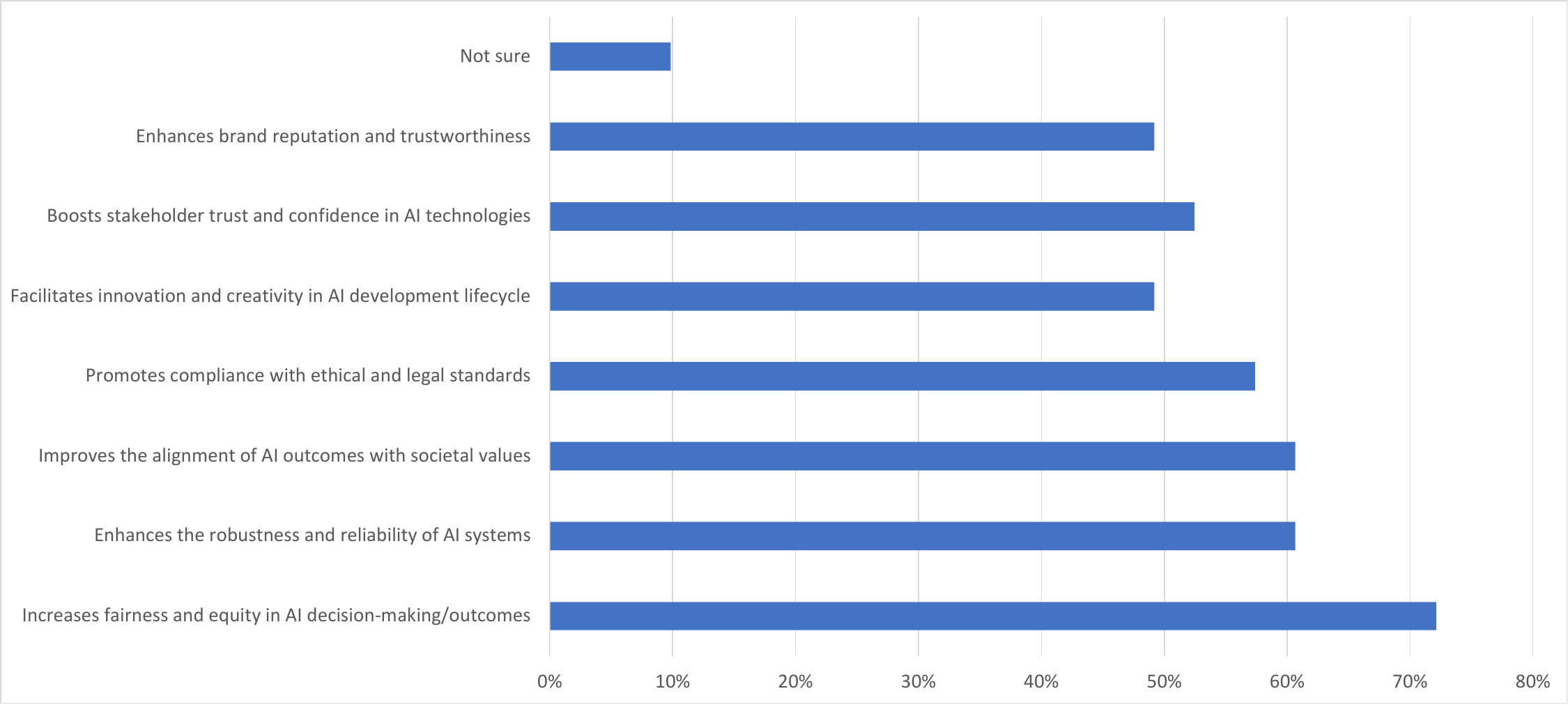}
\caption{Perceived Impact: Effect of D\&I on processes and systems}
\label{Fig:PISP}
\end{figure}

The survey reveals that most respondents acknowledge D\&I's role in enhancing fairness (44\%), strengthening system reliability (37\%), and ensuring compliance with ethical standards (35\%). Additionally, 30\% highlight its impact on stakeholder trust and brand reputation, reinforcing the broader organisational benefits of inclusive AI. Interestingly, early to mid-career professionals (70\%), particularly from large organisations, are the strongest advocates for fairness. In contrast, 83\% of those uncertain about D\&I’s impact are men aged 35–44.

\begin{figure}[hbt!]
\centering
\includegraphics[width=0.8\textwidth]{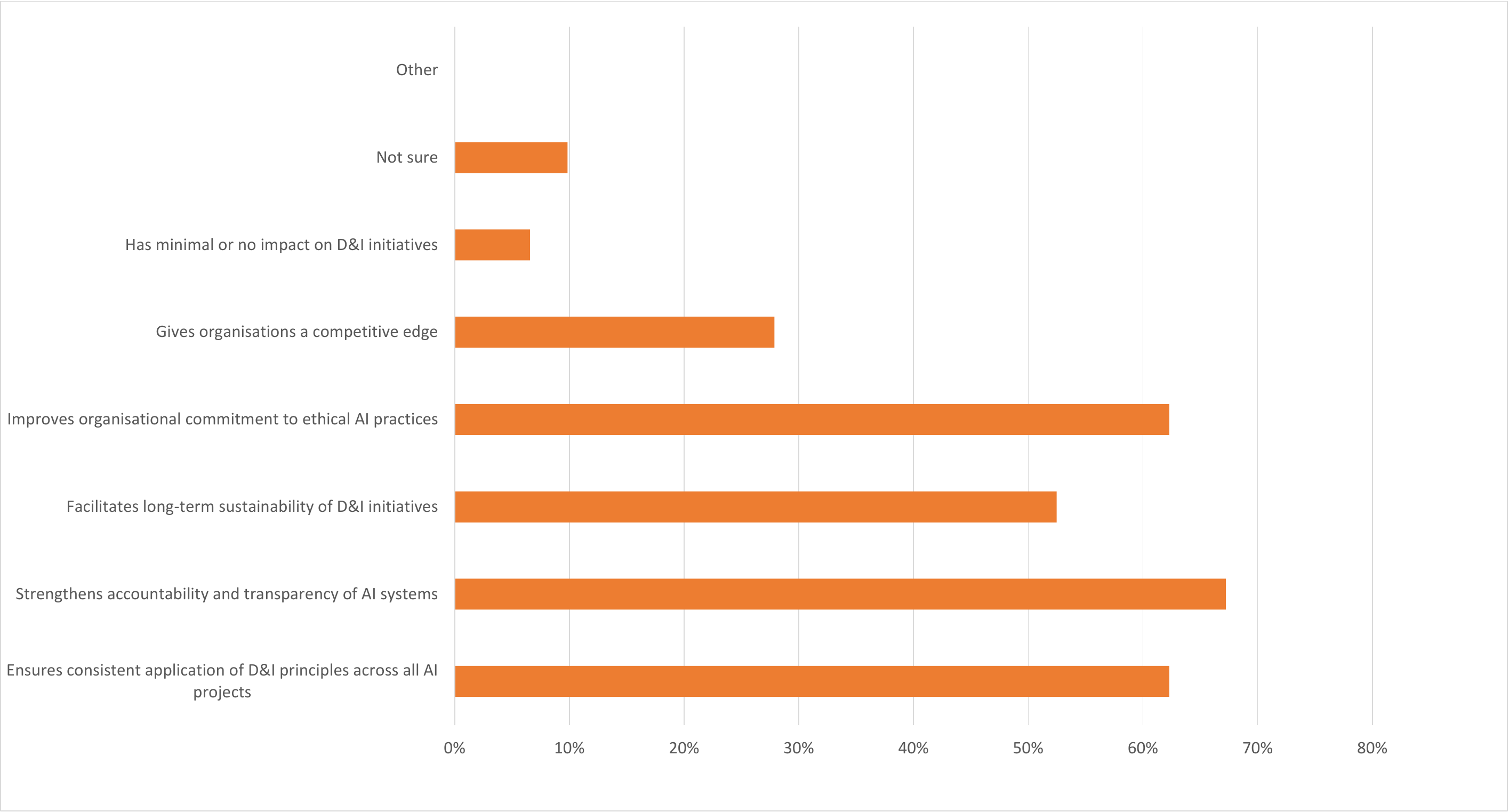}
\caption{Perceived Impact:Effect of government policies on success of D\&I in AI initiatives}
\label{Fig:PIG}
\end{figure}

Governance policies play a significant role in ensuring accountability and consistency in AI diversity and inclusion efforts (see Figure ~\ref{Fig:PIG}. While 67\% of respondents acknowledge governance’s role in ethical AI development and 62\% recognise its importance in organisational commitment to D\&I, only 28\% view governance as a competitive advantage. This suggests that many organisations treat governance as a compliance necessity rather than an enabler for long-term innovation and trust-building. 

\begin{tcolorbox}[highlightstyle]
\textbf{RQ1:} \textit{How do practitioners perceive the influence of D\&I in AI systems?}

It can be observed that reducing biases is a common perceived impact across all five pillars: human, data, systems, processes, and governance. This perceived impact is mostly shared between respondents from larger organisations and those in early to mid-career stages.
\end{tcolorbox}

\subsection{Current D\&I Practices in Organisations}
Diversity and inclusion practices within organisations vary significantly. The most commonly citepd D\&I approach is in active recruitment (51\%), followed by training (44\%), inclusive policies (43\%), and diverse leadership (43\%). However, employee resource groups are less common (31\%), and 23\% of respondents are unsure about their organisation’s D\&I efforts, highlighting ineffective communication. 
Cross-analysis of demographics shows that active recruitment and hiring initiatives are most prevalent in large organisations (48\%) and small firms (32\%), with mid-sized companies lagging behind. Uncertainty about D\&I initiatives is more common among early-career professionals (75\%) and development roles (67\%), particularly in large organisations.

\begin{figure}[hbt!]
\centering
\includegraphics[width=0.8\textwidth]{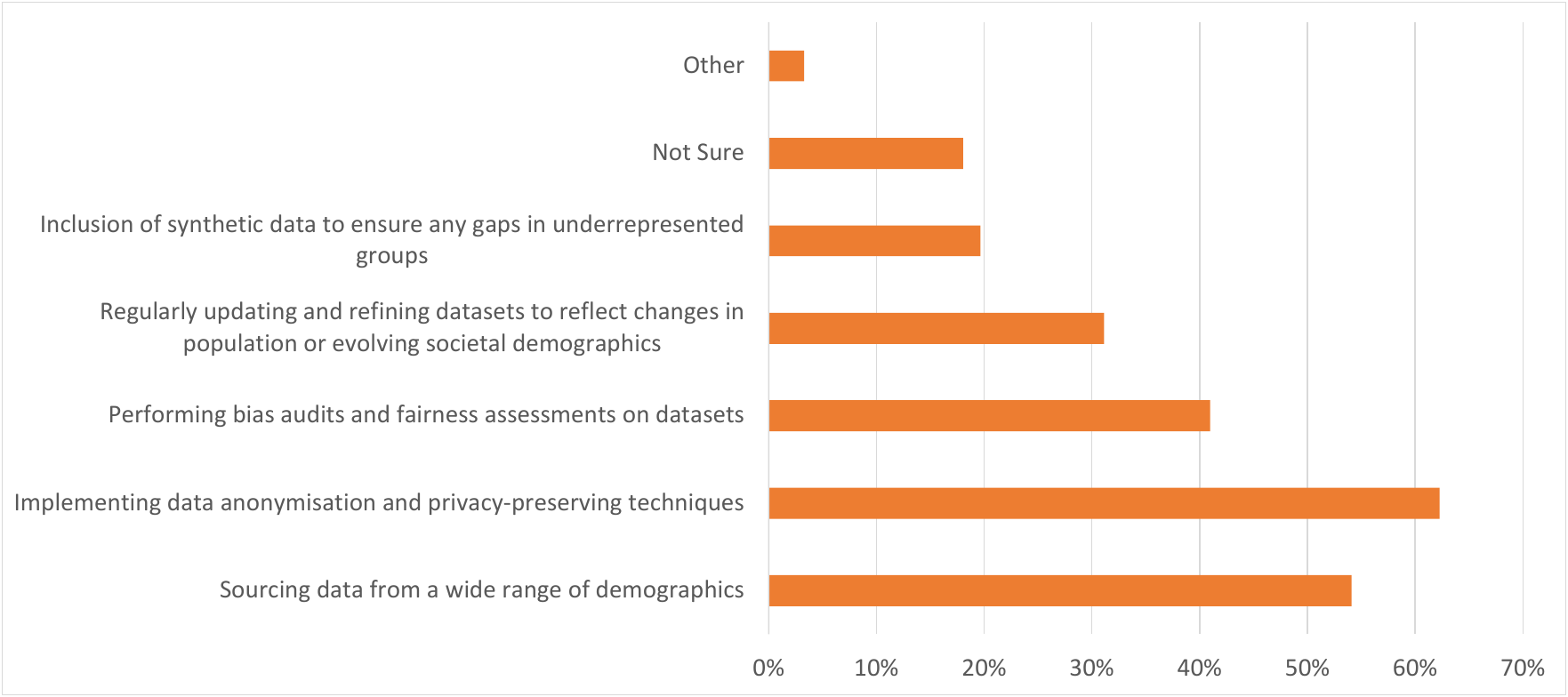}
\caption{Current Practices: Ensuring diverse and inclusive datasets }
\label{Fig:CP-D}
\end{figure}

Figure~\ref{Fig:CP-D} shows that organisations prioritise privacy over fairness in data practices. The most common approach to ensuring diverse and inclusive datasets is data anonymisation (62\%), followed by sourcing data from diverse demographics (54\%). However, bias audits (41\%) and regular dataset updates (31\%) are less frequent, while synthetic data remains relatively underutilised (20\%). Most respondents uncertain about data diversity practices work in large organisations (67\%), where majority are early-career professionals (75\%) indicating a gap between policy and practice.

\begin{figure}[hbt!]
\centering
\includegraphics[width=0.9\textwidth]{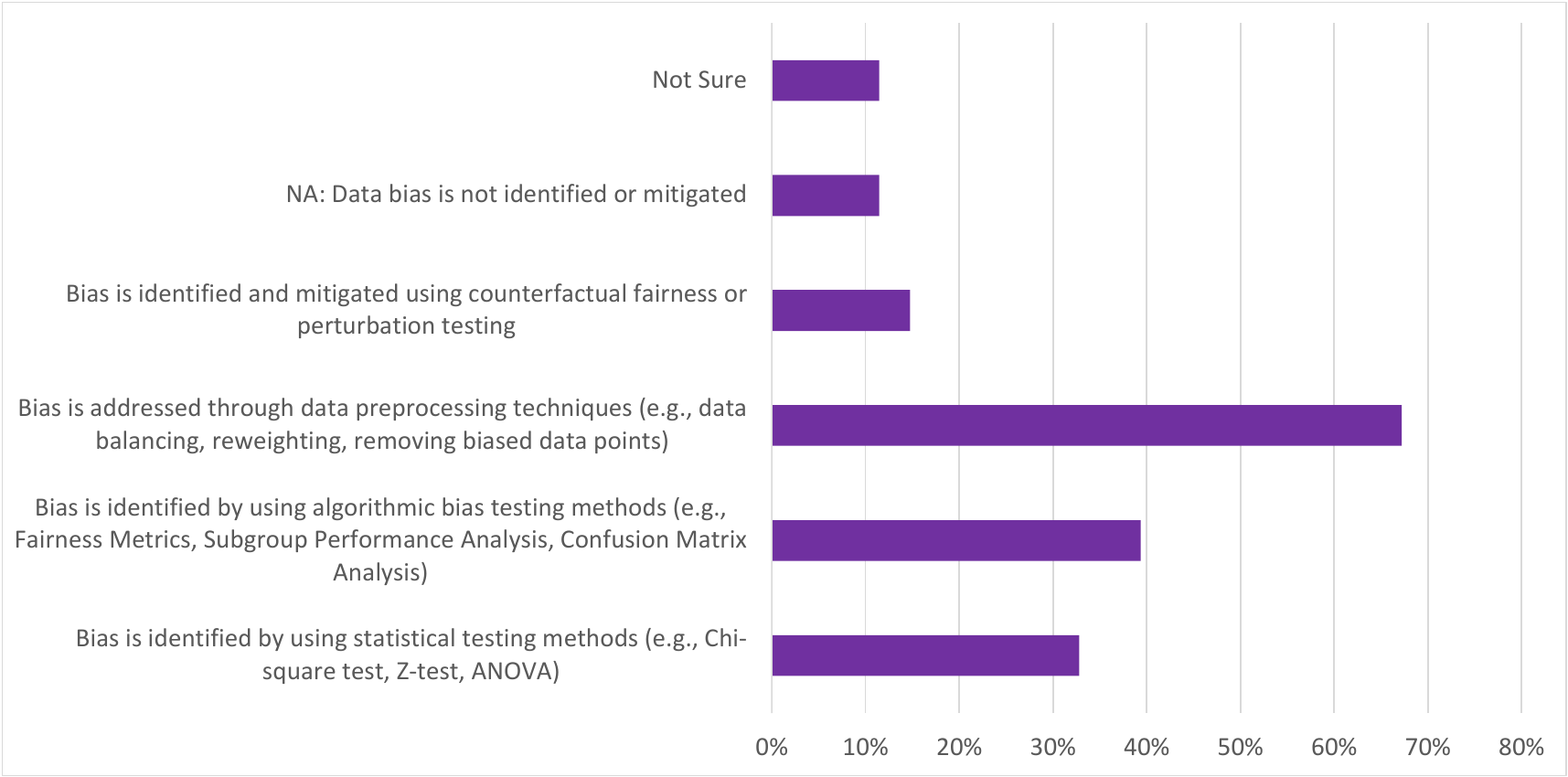}
\caption{Current Practices: Bias mitigation strategies in datasets }
\label{Fig:CP-D2}
\end{figure}

The frequency analysis of bias mitigation strategies (see Figure~\ref{Fig:CP-D2}) reveals that data preprocessing techniques (e.g., data balancing, reweighting) are the most commonly deployed (67\%) followed by Algorithmic bias testing methods (39\%), while statistical testing methods are at 33\% . More advanced techniques, such as counterfactual fairness or perturbation testing, are less frequently utilised (15\%), which indicates sophisticated bias mitigation strategies are not very common. Interestingly, 11\% of respondents either do not identify or mitigate data bias, or uncertain about bias mitigation practices.

We also evaluated the consideration of D\&I principles in AI lifecycle. D\&I principles are incorporated during the pre-development and development stages by 59\% of respondents, reflecting early-stage awareness in AI planning, design, and development. However, only 41\% report D\&I considerations in the post-development phase, indicating limited focus on ongoing monitoring and feedback integration.

15\% of respondents state that D\&I principles are not incorporated in their organisations. Among them, 67\% have mid-senior level experience, 44\% have over six years of experience. Gender wise, 56\% are men and only 22\% are women. In addition, 10\% of respondents are unsure about D\&I incorporation in AI lifescyle.

The most common method for evaluating inclusiveness in AI systems is ongoing monitoring (51\%), followed by implementing feedback loops (46\%). Regular audits (21\%) and external reviews (23\%) are less common. 30\% of respondents are unsure about the evaluation methods of the AI systems. Our cross analysis of  majority (61\%) of those unsure about AI inclusiveness evaluations are early-career professionals, and 56\% work in large organisations.

\begin{figure}[hbt!]
\centering
\includegraphics[width=0.8\textwidth]{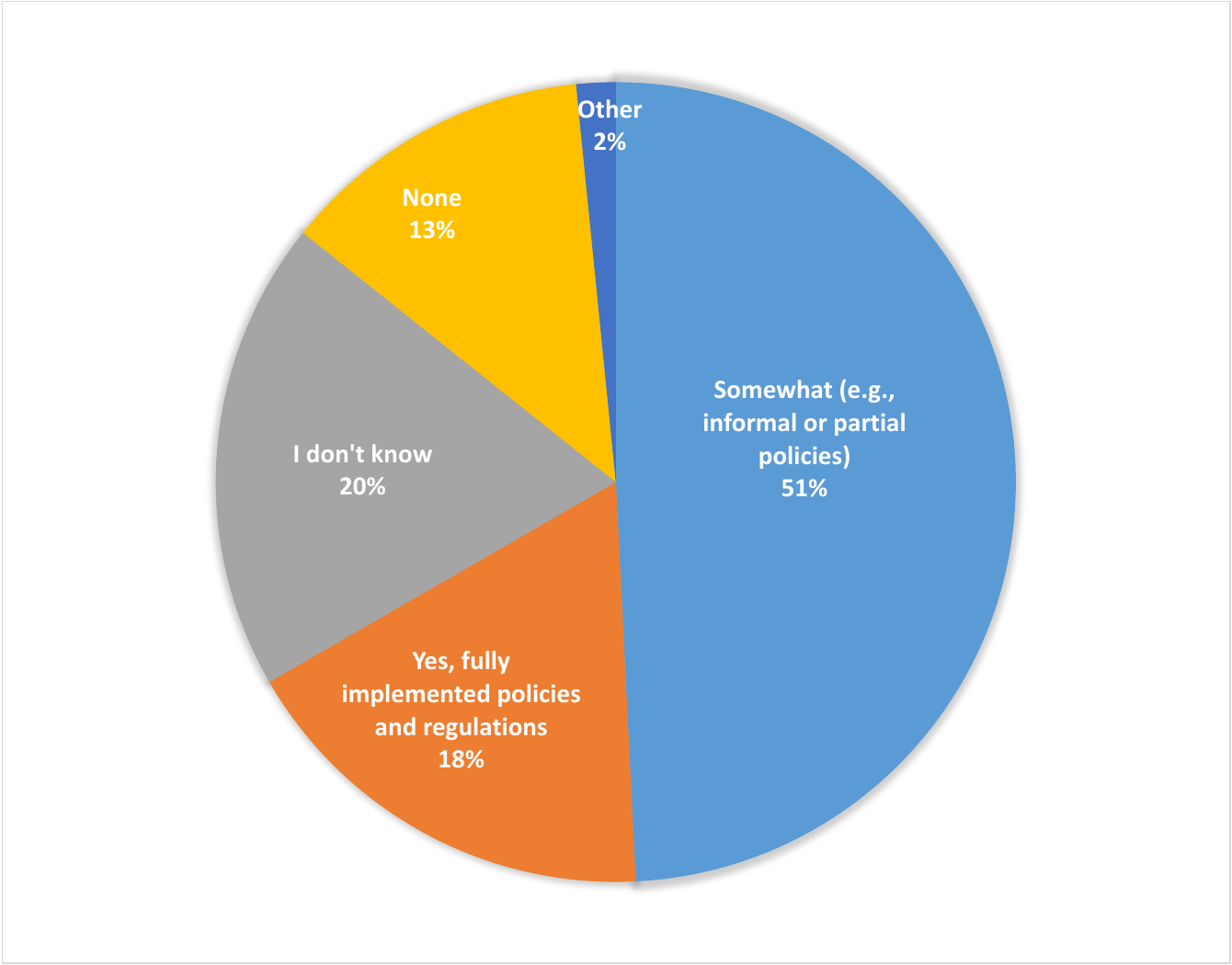}
\caption{Current Practices: Government Structures supporting D\&I }
\label{Fig:CP-G}
\end{figure}

Almost half of respondents (51\%) indicate that their organisations have somewhat established governance structures or policies for D\&I in AI as shown in Figure~\ref{Fig:CP-G}. Only 18\% report fully implemented policies, while 20\% are unsure (I don’t know). Among respondents who do not know, 13\% of the respondents confirm that no such policies exist. Cross analysis of those 13\% with demograhics show that hey work in small to medium-sized organisations (1-200 employees).

\begin{tcolorbox}[highlightstyle]
\textbf{RQ2:} \textit{What strategies and practices are currently implemented by organisations in the AI/ML industry to foster Diversity and Inclusion?}

We found that D\&I practices across AI organisations vary, with active recruitment, training, inclusive policies, and diverse leadership being common but inconsistently applied. Organisations prioritise privacy and data anonymisation, with less emphasis on advanced bias mitigation methods or regular inclusiveness evaluations. Early-career professionals often exhibit uncertainty, indicating a disconnect between organisational policies and practical implementation.
\end{tcolorbox}

\subsection{Challenges in Achieving D\&I in AI}
\begin{figure}[hbt!]
\centering
\includegraphics[width=0.9\textwidth]{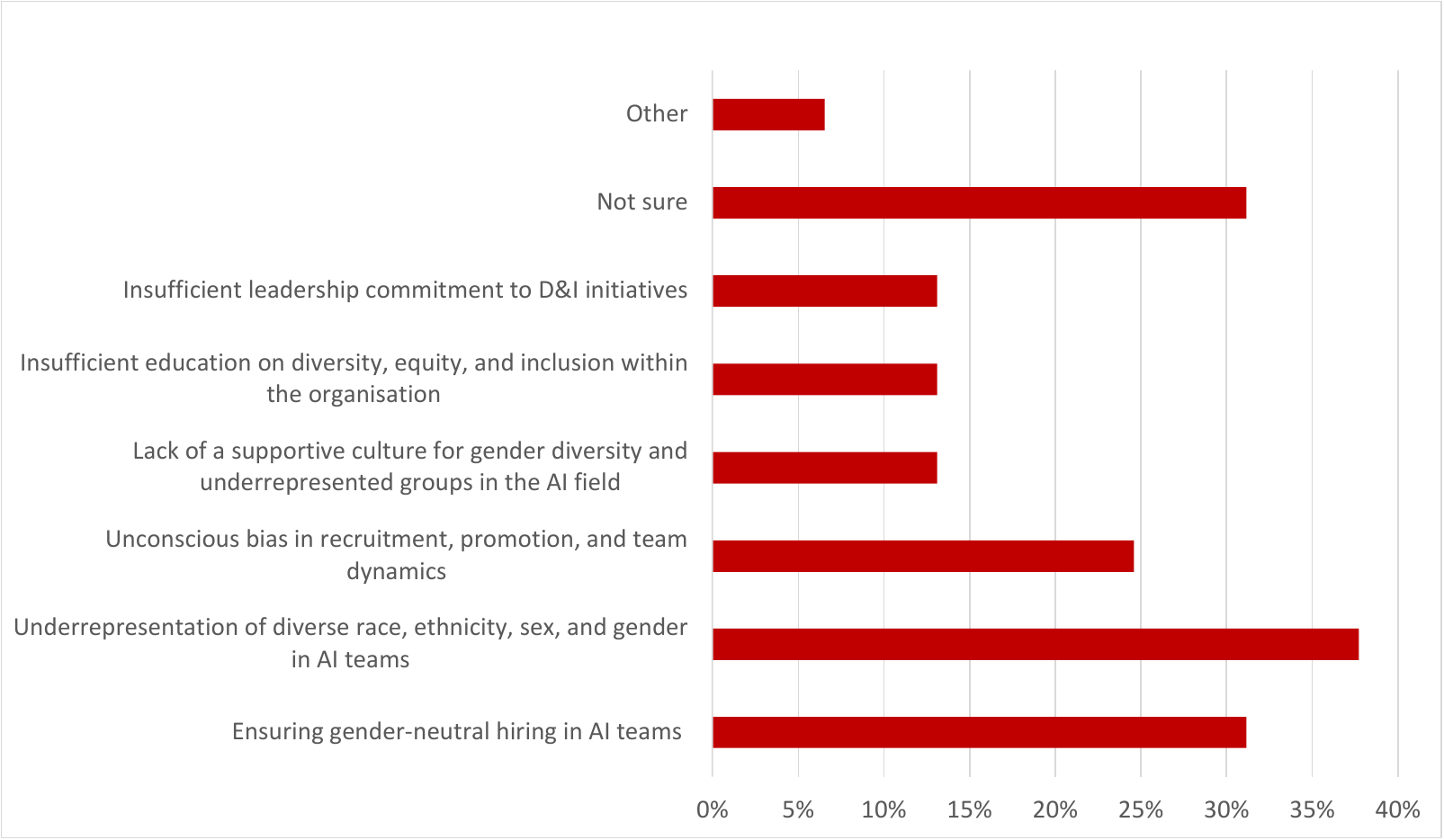}
\caption{Challenges in diverse and inclusive workforce }
\label{Fig:C-H}
\end{figure}

As shown in Figure~\ref{Fig:C-H}, the most common challenges organisations face in achieving a diverse and inclusive AI workforce include the underrepresentation of diverse races, ethnicity, and genders (38\%) and ensuring gender-neutral hiring (31\%). Unconscious bias in recruitment and team dynamics remains a concern for 25\% of respondents. However, a significant portion (31\%) is unsure about the specific challenges. 

Nearly half of the respondents (49\%) identified the lack of comprehensive and accurate demographic data collection as a key challenge in sourcing diverse datasets for AI. Additionally, 46\% pointed to difficulties in recognising and addressing biases due to insufficient knowledge, tools, or data anonymisation techniques. 

Key challenges in integrating D\&I practices into the AI lifecycle include difficulty embedding these principles into existing workflows (25\%), misalignment between AI project goals and organisational D\&I objectives (23\%), and a lack of training and awareness among AI practitioners (22\%).


The most critical challenge in creating inclusive AI systems (see Figure~\ref{Fig:C-sys} in Appendix) is the difficulty in recognising and mitigating unconscious biases (57\%), which significantly impacts fairness and reliability. Ensuring fair representation of intersectional identities (43\%) further complicates inclusivity efforts, as AI systems struggle to account for diverse user experiences. A substantial lack of trust in AI software’s accuracy and fairness (39\%) suggests broader concerns beyond technical limitations. Additionally, insufficient monitoring and auditing processes (38\%) indicate gaps in ongoing bias assessment.

The biggest challenge in AI governance is the lack of clear D\&I guidelines (54\%), leading to inconsistent enforcement across teams (41\%). Difficulty in measuring the impact of D\&I policies (33\%) and aligning them with organisational goals (30\%) further complicates implementation. While stakeholder resistance is relatively low (13\%), accountability and transparency remain a concern for some (25\%).

\begin{tcolorbox}[highlightstyle]
\textbf{RQ3:} \textit{What are the primary challenges faced by AI/ML organisations in establishing and sustaining Diversity and Inclusion in AI?}, 

Organisations face key challenges which include under-representation of diverse groups, insufficient collection of demographic data, and difficulty integrating D\&I principles into existing AI workflows. Organisations also struggle to identify subconscious biases, ensure intersectional fairness in AI systems, and consistently enforce clear governance guidelines throughout the AI lifecycle.
\end{tcolorbox}

\begin{figure}[hbt!]
\centering
\includegraphics[width=0.8\textwidth]{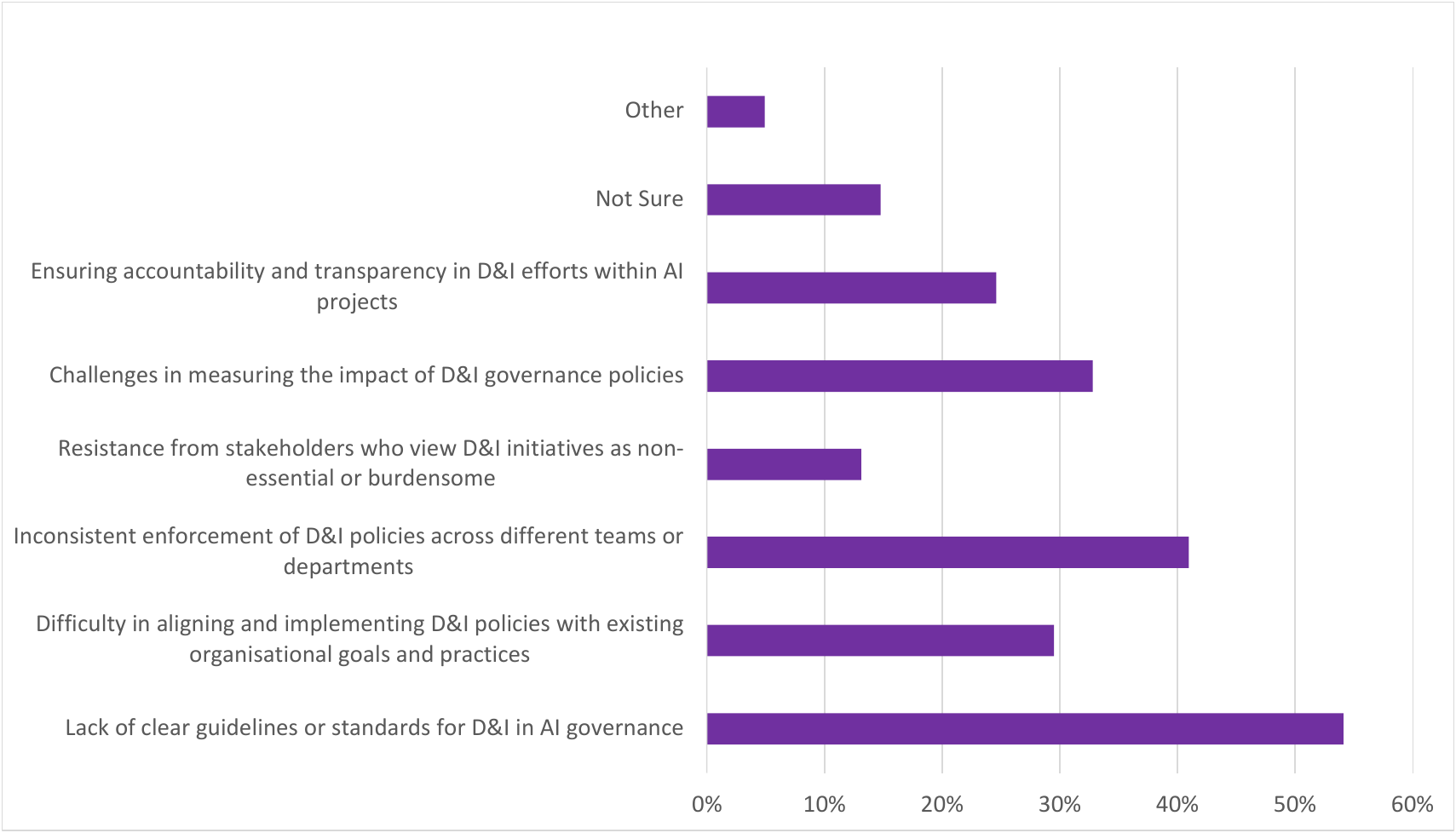}
\caption{Challenges in establishing D\&I governance policies }
\label{Fig:C-sys}
\end{figure}

\section{Discussion}\label{sec:discussion}
This section explores the disconnect between the acknowledged importance of D\&I in AI and its inconsistent implementation across organisational practices, and the persistance challenges faced throughout the development processes, governance frameworks, and workforce dynamics.

\subsection{Link Between Perceived Impact, Current Practices, and Challenges of D\&I in AI}
The survey results reveal a significant disconnect between the perceived importance of D\&I in AI and its actual implementation across organisations. While a majority of respondents (67\%) believe that diverse teams help reduce bias in AI systems, this belief is not consistently reflected in hiring, promotion, and governance structures. Similarly, although 87\% of respondents recognise that diverse datasets improve AI fairness, only 41\% report that their organisations conduct bias audits to evaluate dataset inclusiveness. This gap highlights the barriers between perception and execution of D\&I in AI. The following sections further summarise our discussion with respect to the RQ1, RQ2 and RQ3 across the five pillars of the AI ecosystem. 

\subsubsection{Perceived Impacts}
The findings reveal nuanced perceptions about the perceived impact of D\&I across the five pillars of the AI ecosystem. Reducing biases is a common perceived impact across all five pillars. However, scepticism was noted regarding the impact of D\&I on diverse teams, AI systems, and governance policies. 10\% of the respondents reported minimal or no real impact of diverse teams. Most of these respondents are developers who may perceive bias as a purely technical issue that can be mitigated through algorithmic bias mitigation techniques, and they may fail to realise  the broader benefits of diverse teams. Such perceptions resonate with the work of Ferrara~\citep{ferrara2023fairness} and Mehrabi et al.~\citep{mehrabi2019survey}, who highlight how technical approaches to fairness may overshadow more holistic and systemic diversity considerations. Furthermore, scepticism was notably higher within highly regulated industries, such as healthcare and finance, where compliance, privacy, and technical constraints potentially overshadow or complicate the integration of D\&I  in AI practices. While governance policies are widely seen as essential for transparency and accountability in AI systems~\citep{rakova2021where, wef2022blueprint}, some managers (particularly in smaller organisations) perceive limited impact. This may reflect their broad responsibilities that can constrain deeper engagement with D\&I governance strategies.

\subsubsection{Current Practices}

One key insight is the decline in D\&I consideration from the development phase (59\%) to post-development (41\%), suggesting a lack of ongoing oversight. This trend is especially noticeable in larger organisations (67\%), where formal policies exist but are not consistently applied. Many early-career professionals also reported uncertainty about their organisation’s D\&I in AI efforts, pointing to possible communication and visibility gaps. These findings align with existing research that highlights the importance of continuous engagement, clear communication, and accountability~\citep{rakova2021where, shams2023ai, wef2022blueprint}. Another observation is the reliance on data anonymisation (62\%) as a primary bias mitigation method. While effective for privacy, it may also obscure underlying biases, making them harder to detect and address proactively.

\subsubsection{Challenges}
The key challenges in integrating D\&I practices into the AI lifecycle (according to the five pillars include lack of comprehensive and accurate demographic data (49\%), difficulty
in embedding D\&I in AI principles into existing workflows (25\%),  difficulty in recognising and mitigating subconscious biases (57\%), and  the lack of clear D\&I in AI guidelines (54\%). 
An interesting finding is that data anonymisation emerged both as the most widely adopted bias mitigation method and as a notable challenge for recognising biases in datasets. This highlights a challenge for organisations struggling between maintaining privacy and ensuring fairness. Additionally, many respondents unsure about challenges in sourcing diverse data were early-career professionals. This suggests that when data governance is handled by specialised teams, awareness of D\&I considerations among technical staff may be unintentionally limited.
This mixed perception of data anonymisation supports earlier studies that discuss the challenge  between protecting privacy and effectively recognising biases in data~\citep{ferrara2023fairness, mehrabi2019survey}. While anonymisation safeguards privacy, it often hides demographic details needed to evaluate and correct biases, complicating fairness efforts~\citep{obermeyer2019dissecting}. Furthermore, limited awareness among early-career professionals highlights organisational challenges, where specialised teams handling data governance might restrict visibility into D\&I in AI practices~\citep{rakova2021where}. 
The challenges we identify in this study closely align with the challenges identified by Shams et al.~\citep{shams2023ai}, particularly,  the under representation of diverse demographics and the presence of unconscious biases affecting recruitment and team dynamics. However, our findings reveal some additional challenges. A significant portion of respondents (31\%) expressed uncertainty about specific D\&I in AI challenges. This indicates a potential gap in D\&I awareness or learning among AI practitioners. Furthermore, we identified difficulties in embedding D\&I principles into existing workflows (25\%) and aligning them with organisational goals (23\%), suggesting that operational integration of D\&I initiatives remains a substantial hurdle. These insights highlight the evolving nature of D\&I challenges in AI and calls for continuous adaptation and awareness of robust strategies to effectively integrate D\&I principles throughout the AI lifecycle. 

\subsection{Perception vs. Reality in Workforce Diversity}
Our survey reveals a clear gap between practitioners’ recognition of the value of D\&I and actual organisational practices. 70\% of early-career professionals (1–5 years of experience) believe that diverse teams improve AI fairness, yet hiring practices do not reflect this belief. Women and ethnic minorities face greater challenges in recruitment and promotion, with only 26\% of women in our survey citing barrier to career progression, compared to 70\% of men. Gender-neutral hiring is also perceived as a challenge by 31\% of the respondents.
Despite this, only 50\% of the respondents confirm their organisations have active recruitment strategies to increase workforce diversity. Some (20\%) of respondents are unsure whether their organisation has a formal D\&I hiring policy, indicating a lack of transparency in workforce diversity efforts. Furthermore, the leadership composition in AI remains heavily male-dominated, with 67\% of development roles, 68\% of management roles, and 66\% of governance roles occupied by men.
The gap between perception and practice is obvious, as many professionals acknowledge the importance of D\&I in AI, recruitment and retention policies are not fully aligned with this recognition.

\subsection{Governance vs. Algorithmic Fairness}
Governance policies in AI are often seen as a key mechanism for ensuring ethical AI development, yet technical professionals remain sceptical about their effectiveness. Results show that while 45\% of respondents believe governance structures improve AI accountability and transparency of AI systems, only 18\% of the managers agree that governance policies have a direct impact on AI fairness.
Despite these perceptions, 55\% of respondents reported some presence of structured D\&I governance policies, whereas 70\% of respondents from small companies (1–50 employees) stated that no such policies exist.
A further issue lies in the methods used for bias mitigation. Organisations prioritise data anonymisation (62\%) over proactive fairness strategies such as bias audits (41\%) or algorithmic fairness testing (only 15\%). This suggests that organisations focus more on compliance-driven approaches rather than actively embedding fairness into AI systems. Additionally, our results show that D\&I considerations decline after the initial development phase. This raises concerns about whether governance structures are truly integrated into AI development processes or simply exist for regulatory compliance purposes.

\subsection{Recognition vs. Implementation of Data Diversity}
Our survey results show widespread agreement that diverse datasets are critical for AI fairness, with 87\% of respondents stating that data diversity helps reduce bias, and 70\% believing that it improves AI’s ability to generalise across populations. Despite this recognition, data diversity challenges remain unresolved in many organisations.
Only 41\% of the respondents believe that regular bias audits on their datasets can ensure D\&I, while 18\% of respondents are unsure about the effect of data on D\&I in AI. This suggests that although there is broad recognition of the importance of data diversity, there remains a lack of clarity or consensus on how to implement effective auditing and evaluation processes. While our findings show that scepticism about data diversity exists even within regulated industries such as healthcare and finance, this may reflect the complexity of compliance and ethical constraints in such sectors. In contrast, the lower frequency of bias audits and advanced fairness checks observed across the sample could suggest that in technology-focused roles, performance optimisation may take precedence over dataset inclusiveness. A key takeaway from this finding is that organisations recognise the importance of diverse datasets but do not apply standardised practices to ensure data inclusiveness throughout the AI system lifecycle.

\subsection{Ethnicity and its Impact on AI Diversity}
Our survey shows that 70\% of respondents identified as Asian, with limited representation from other ethnic groups such as Caucasian (13\%), Middle Eastern or North African (10\%), and minimal respondents identifying as Indigenous, Black, or Southern European. Among women respondents, 13 identified as Asian and only 4 as Caucasian, indicating that ethnic diversity within gender minorities remains low.

While technical roles in AI in many parts of the world are increasingly filled by professionals from Asian and migrant backgrounds, leadership and governance roles remain disproportionately held by men, especially from Western contexts. This is reflected in our findings where men occupy 70–74\% of development, management, and governance roles. Women, particularly from underrepresented ethnic backgrounds, reported more career barriers.

While 67\% of respondents believe that diverse teams reduce bias, the lack of ethnic representation in both technical and leadership roles suggests this belief is not yet realised in practice. The under representation of ethnic and gender minorities in leadership limits the breadth of perspectives shaping AI governance and ethics. Ensuring D\&I at decision making levels is essential for diverse and inclusive AI development.

\subsection{Current Trends against D\&I and their Impact}
Between late 2024 and 2025, D\&I efforts in the AI sector have undeniably been rolled back at multiple levels. Government actions in the US and UK have removed incentives or introduced disincentives for pursuing D\&I practices, framing it as a distraction from innovation or even a legal liability. In parallel, many AI-driven companies and tech giants have scaled down or re-branded their DEI and responsible AI programs, often under economic or political pressures. These developments are interrelated with a broader backlash against ``woke" tech, a narrative that paints efforts to make AI fairer as ideologically suspect. The impacts of this retreat are far reaching: from less diverse engineering teams and weaker fairness checks in AI products to shifts in AI governance that place less emphasis on justice and inclusion. These trends validate the concerns raised in our survey results, such as the gap between D\&I perception and practice, declining oversight of AI impacts post-deployment, and low confidence in D\&I among practitioners. If not addressed, the rollback of D\&I threatens to exacerbate biases in AI systems and erode trust in AI’s role in society. The cutbacks in D\&I in AI are alarming because without it, the technologies shaping our future may be less safe, less accurate, and less reliable \footnote{\href{https://medium.com/@jasminemoradi/the-cut-of-dei-in-tech-threatens-inclusive-ai-future-10-strategies-for-how-tech-companies-can-ac08676c5ad8}{The Cut of DEI in Tech Threatens Inclusive AI Future}}.

\section{Threats to Validity}
As with any survey-based research, this study is subject to several limitations that may affect the interpretation and generalisability of the findings. We outline key considerations regarding internal, external, construct, and methodological validity.

\subsection{Internal Validity}
Internal validity refers to the extent to which the findings accurately reflect the genuine views and experiences of the participants. One potential concern is self-selection bias. Participation in the survey was voluntary, which may have attracted individuals with a particular interest in D\&I in AI or stronger opinions on the topic. This could skew the results toward more favourable or critical perspectives. Although the survey was pilot-tested to ensure clarity and relevance, self-reported data always carry the risk of recall inaccuracies and social desirability bias. Respondents may have unintentionally misrepresented their organisation’s practices, either positively or negatively, based on perception rather than objective knowledge. These risks were mitigated through careful wording and by ensuring anonymity, but they cannot be entirely eliminated.

\subsection{External Validity}
We received 61 valid responses, offering valuable insights but limiting the statistical generalisability of the results. The sample was not demographically balanced, with an over-representation of participants identifying as Asian and a majority identifying as male. Participation from other groups, particularly Indigenous, Black, and minority ethnic women, was minimal. As a result, the findings may not capture the full range of experiences across different cultural or organisational contexts. While the data reflect meaningful patterns within the respondent group, caution should be taken when extending these findings to the broader AI workforce.

\subsection{Construct Validity}
This study aimed to explore perceptions, practices, and challenges related to D\&I in AI organisations. These are inherently complex and context-dependent concepts. While the survey included definitions and examples to guide interpretation, respondents may still have held varying understandings of key terms such as ‘diversity’, ‘inclusion’, and ‘bias mitigation’. The use of multiple-choice questions, although necessary for analysis, may have restricted the expression of more nuanced perspectives. Open-ended questions were included to allow participants to elaborate on their views, but the depth of insight is still limited compared to qualitative methods such as interviews or focus groups.

\subsection{Methodological Constraints}
This study was cross-sectional, capturing a snapshot of participants’ perceptions at a single point in time. It does not account for how these views may shift in response to organisational changes, broader industry trends, or evolving public discourse. The use of a single survey instrument, disseminated through specific networks, may also have influenced who was reached and who chose to participate. Although the study provides useful insight into D\&I from the perspective of a specific cohort, future research could strengthen these findings through longitudinal studies, larger and more diverse samples, and complementary qualitative methods.

Despite these limitations, the findings offer meaningful insights into how professionals in the AI and ML sector perceive and experience D\&I within their organisations. The results reflect emerging patterns around workforce composition, perceived benefits of diverse teams, and the practical challenges in implementing D\&I practices. These insights can inform future policy development, organisational strategy, and research by highlighting where current efforts are effective and where gaps remain. In particular, the perspectives of early-career professionals and respondents from underrepresented backgrounds provide important direction for improving inclusivity in AI development and governance.

\section{Conclusion and Future Work}
\label{sec:Conclusion}
As AI systems increasingly shape critical aspects of society, embedding D\&I into their development is no longer optional; it is foundational to fairness, trust, and innovation. This study surveyed AI practitioners to examine how D\&I principles are perceived, practised, and challenged across the industry, addressing three key research questions on impacts, practices, and barriers. While the findings highlight widespread agreement on the value of D\&I, particularly in reducing bias and enhancing creativity, they also expose a persistent gap between intention and execution. Organisations often fall short in translating ethical commitments into concrete actions, particularly in post-development stages and governance. Challenges such as under-representation of marginalised groups, insufficient demographic data, and limited integration of D\&I into workflows remain pressing.

Future research should focus on developing deeper empirical evidence and scalable methodologies to evaluate the long-term impact of D\&I initiatives in AI. \textit{For researchers}, the next step is to build robust, longitudinal studies that quantify the tangible impacts of D\&I initiatives. There's a need for interdisciplinary work that bridges algorithmic fairness with social and organisational dynamics, especially across industries and regions. \textit{For practitioners}, embedding D\&I into everyday workflows requires actionable tools, regular audits, and team-wide training tailored to different organisational roles. Efforts should align with business objectives to ensure inclusive development is not peripheral but integral. \textit{For policymakers and governance-setting bodies}, the mandate is clear: introduce enforceable regulations, transparency standards, and incentives that elevate D\&I from aspiration to norm. Certification schemes and public accountability mechanisms can catalyse industry-wide adoption.

Ultimately, embedding D\&I practices into AI system development and deployment serves as both an ethical imperative and a strategic advantage, fostering innovation, enhancing organisational resilience, and strengthening societal trust in AI systems. This study calls upon industry stakeholders, policymakers, and researchers to collaborate proactively in driving measurable and sustainable progress towards genuinely inclusive AI ecosystems.


\bibliographystyle{apalike}
\bibliography{bibliography}
\newpage


\section*{Supplementary material}

# Appendix A: D&I in AI Survey Instrument

## Section: Demographics

Help us understand the diverse backgrounds of participants in the AI ecosystem by sharing some general information about yourself. Your responses will remain anonymous.

**Q1: Which age group best represents you?**
- Under 25
- 25-34
- 35-44
- 45-54
- 55 and older
- Prefer not to say

**Q2: What gender do you identify with?**
- Male
- Female
- Non-binary/Third gender
- Prefer not to say

**Q3: What ethnicity best represents your background?**
- Indigenous
- Caucasian
- Black or African Descent
- Asian
- Middle Eastern or North African
- Hispanic or Latino
- Please mention any other option not listed above in "Other" below: _____

**Q4: What is your current role in the AI/ML industry?**
- Development: Team leaders, Architects, Developers, UX/UI designers, Data scientists, Testers, Business Analysts and Operators
- Management: Board Member, Executives, Managers Trainer/Educator
- Governance: Responsible AI governors, AI technology producers, AI policy advisors
- Prefer not to say
- Other (Please specify): _____

**Q5: How long have you been working in the AI/ML industry?**
- Less than 1 year
- 1-5 years
- 6-10 years
- More than 10 years

**Q6: How many employees does your organisation have?**
- 1-50
- 51-200

- 201-1000
- More than 1000

**Q7: What types of AI systems have you worked on or are currently working on?**

- Recognition Systems (e.g., Facial Recognition, Speech/Voice Recognition)
- Recommender Systems
- Predictive Models/Analytics
- Image Processing/Computer Vision
- Large Language Models (LLMs) (e.g., GPT, BERT, LLaMA)
- Generative AI (e.g., DALL-E, MidJourney)
- Classification Systems
- Please provide any option not listed above in "Other" below: _____

**Q8: What AI application domains have you worked in or currently are working in?**
- Human Resource
- Banking and Finance
- Insurance
- Health and Medical
- Defence
- Education
- Science and Research
- Infrastructure
- Energy
- Entertainment/Gaming

**Q9: Do you wish to share any other demographic characteristics that might be relevant?**
- No
- Yes (Please mention in Other below)
- Other (please specify): _____

# Section: Current Practices
The questions in this section are related to the **current strategies and practices** used by your organisation to embed or integrate Diversity and Inclusion (D&I) principles in AI.

### Humans

**Q10: What practices are in place in your organisation to ensure diverse and inclusive teams are working on AI design, development, and deployment? (Select all that apply)**

- Active Recruitment and hiring practices

- D&I Training for employees
- Organisational policies for inclusive teams
- Diverse Leadership and decision-making committees
- Creating diversity-focused employee resource groups
- Not Sure
- Other (please specify): _____

### Data

**Q11: What methods are used to ensure the datasets for training AI are diverse and inclusive? (Select all that apply)**

- Sourcing data from wide range of demographics
- Implementing data anonymisation and privacy-preserving techniques
- Performing bias audits and fairness assessments on datasets
- Regularly updating and refining datasets to reflect changes in population or evolving societal demographics
- Inclusion of synthetic data to ensure any gaps in underrepresented groups
- Not Sure
- Other methods (please specify): _____

**Q12: How are potential data biases identified and mitigated in your AI development lifecycle? (Select all that apply)**

- Bias is identified using statistical testing methods (e.g., Chi-square test, Z-test, ANOVA)
- Bias is identified using algorithmic bias testing methods (e.g., Fairness Metrics, Subgroup Performance Analysis, Confusion Matrix Analysis)
- Bias is addressed through data preprocessing techniques (e.g., data balancing, reweighting, removing biased data points)
- Bias is identified and mitigated using counterfactual fairness or perturbation testing
- NA: Data bias is not identified or mitigated
- Not Sure
- Other methods of bias identification and mitigation (please specify): _____

### Processes

**Q13: At which stages of the AI development lifecycle are D&I principles incorporated? (Select all that apply)**

- Pre-development (e.g., during initial planning, team formation, and data collection)
- Development (e.g., during model design, algorithm development, and testing)

- Post-development (e.g., during deployment, monitoring, and feedback analysis)
- D&I principles are not incorporated
- Not Sure
- Please add any additional options in Other below: _____

## Systems

**Q14: How are AI systems evaluated for inclusiveness? (Select all that apply)**

- Regular auditing (e.g., conducting bias audits and fairness assessments)
- Ongoing monitoring (e.g., tracking system performance across different demographic groups)
- Implementing a feedback loop (e.g., collecting and analysing user feedback to identify issues related to diversity and inclusion)
- Periodic external reviews (e.g., involving experts or diverse focus groups for usability testing)
- Not Sure
- Other (please specify): _____

## Governance

**Q15: Are there any governance structures, policies, and regulations to support D&I in AI within your organisation? (Select all that apply)**

- Somewhat (e.g., informal or partial policies)
- Yes, fully implemented policies and regulations
- I don't know
- None
- Other (please specify):_____

**Q16: Following up from the above question, if there is an organisational policy, is it inspired by any of the following existing ethical standards?**

- US AI Bill of Rights
- Australian AI Ethics Principles
- EU Guidelines on Trustworthy AI
- Singapore Model AI Governance Framework
- China's AI Ethics Guidelines
- Google's Diversity, Equity, and Inclusion (DEI) Framework
- Microsoft Diversity and Inclusion Strategy
- None of the above

# Section: Challenges

Help us understand the key challenges you or your organisation face in achieving D&I in AI.

## Humans

**Q17: What are the main challenges your organisation has faced in achieving a diverse and inclusive workforce in AI? (Select all that apply)**

- Ensuring gender-neutral hiring in AI teams
- Underrepresentation of diverse race, ethnicity, sex, and gender in AI teams
- Unconscious bias in recruitment, promotion, and team dynamics
- Lack of a supportive culture for women and underrepresented groups in the AI field
- Insufficient education on diversity, equity, and inclusion within the organisation
- Scarcity of women and diverse role models in the AI ecosystem
- Insufficient leadership commitment to D&I initiatives
- Not sure
- Other (please specify): _____

## Data

**Q18: What challenges exist in sourcing and using diverse datasets for AI? (Select all that apply)**

- Inequalities and implicit biases in data related to race, ethnicity, age, and gender
- Lack of comprehensive and accurate demographic data collection
- Difficulty recognizing and addressing biases in data sources due to insufficient knowledge or tools
- Not Sure
- Other challenges (please specify): _____

## Processes:

**Q19: What challenges exist in integrating D&I practices into AI design, development and deployment lifecycles? (Select all that apply)**

- Gender and racial discrimination within the AI design/development team
- Difficulty incorporating D&I practices into existing development workflows and methodologies
- Resistance to adopting D&I practices due to perceived impact on efficiency or performance
- Insufficient D&I training and awareness among AI practitioners
- Challenges in aligning AI project goals with organisational D&I objectives
- Limited tools and frameworks for embedding D&I practices throughout the AI lifecycle
- Not Sure
- Other (please specify): _____

### Systems:

**Q: What challenges are faced in creating diverse and inclusive AI systems?**

- Absence of explicit gender classifiers in AI systems
- Lack of trust in the accuracy and fairness of AI-enabled software
- Difficulty in recognising and mitigating subconscious biases embedded in AI systems
- Challenges in ensuring that AI systems fairly represent intersectional identities (e.g., race, gender, disability)
- Insufficient monitoring and auditing processes to continuously assess and address biases in AI systems
- Not Sure
- Other (please specify): _____

### Governance:

**Q21: What challenges arise in establishing and enforcing D&I in AI governance policies? (Select all that apply)**

- Lack of clear guidelines or standards for D&I in AI governance
- Difficulty in aligning D&I policies with existing organisational goals and practices
- Inconsistent enforcement of D&I policies across different teams or departments
- Resistance from stakeholders who view D&I initiatives as non-essential or burdensome
- Challenges in measuring the impact of D&I governance policies
- Ensuring accountability and transparency in D&I efforts within AI projects
- Not Sure
- Other (please specify): _____

## Section: Perceived Impact
Share your perspectives on how D&I efforts can influence AI systems.

### Humans

**Q22: How do you perceive the impact of having a diverse and inclusive team on the development of a fair AI system? (Select all that apply)**

- Reduces bias in AI outcomes
- Enhances creativity and innovation in AI development
- Leads to more comprehensive problem-solving approaches
- Strengthens user trust and acceptance of the AI system
- Has minimal or no impact on the ethical outcomes of the AI system
- Not sure
- Other (please specify): _____

### Data:

**Q23: In your opinion, how does having diverse and inclusive data impact the ethical outcomes of the AI system?**

- Leads to more accurate and representative AI models
- Reduces the likelihood of biased or discriminatory outcomes
- Improves the AI system's ability to generalise across different populations
- Has minimal or no impact on the AI system's performance
- Not sure
- Other (please specify): _____

### System and Processes

**Q24: What are the perceived benefits of integrating D&I practices into AI systems and processes? (Select all that apply)**

- Increases fairness and equity in AI decision-making/outcomes
- Enhances the robustness and reliability of AI systems
- Improves the alignment of AI outcomes with societal values
- Promotes compliance with ethical and legal standards
- Facilitates innovation and creativity in AI development lifecycle
- Boosts stakeholder trust and confidence in AI technologies
- Enhances brand reputation and trustworthiness
- Not sure
- Other (please specify): _____

### Governance:

**Q25: How do governance policies and structures affect the overall success and sustainability of D&I initiatives in AI?**

- Ensures consistent application of D&I principles across all AI projects
- Strengthens accountability and transparency of AI systems
- Facilitates long-term sustainability of D&I initiatives
- Improves organisational commitment to ethical AI practices
- Gives organisations competitive edge
- Has minimal or no impact on D&I initiatives
- Not sure
- Other (please specify): _____

**Q26: Would you like to make any further comments or recommendations regarding D&I in AI within an industrial context?**

# Appendix B: Additional Figures

Figure 10: Demographics

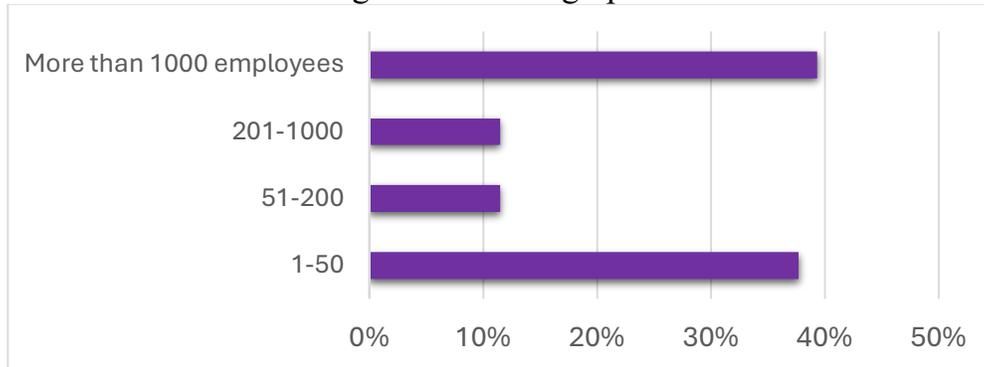

(a) Organisation Size

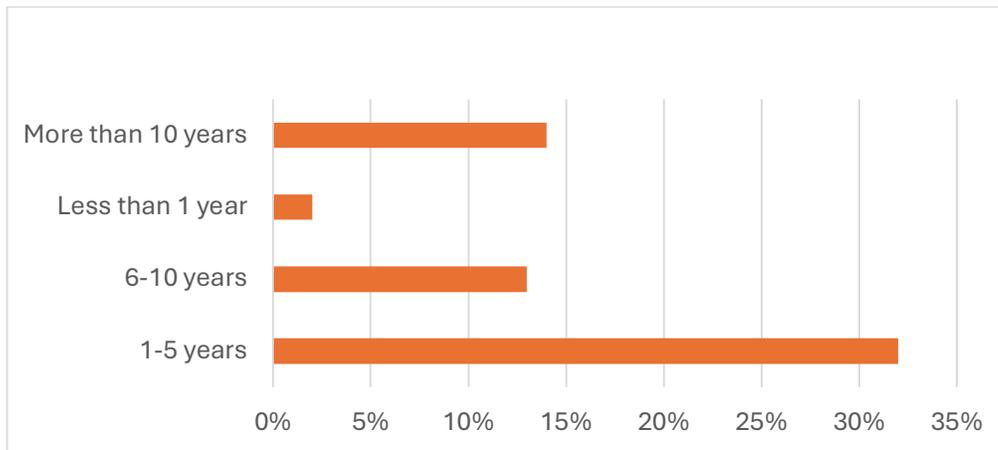

(b) Experience

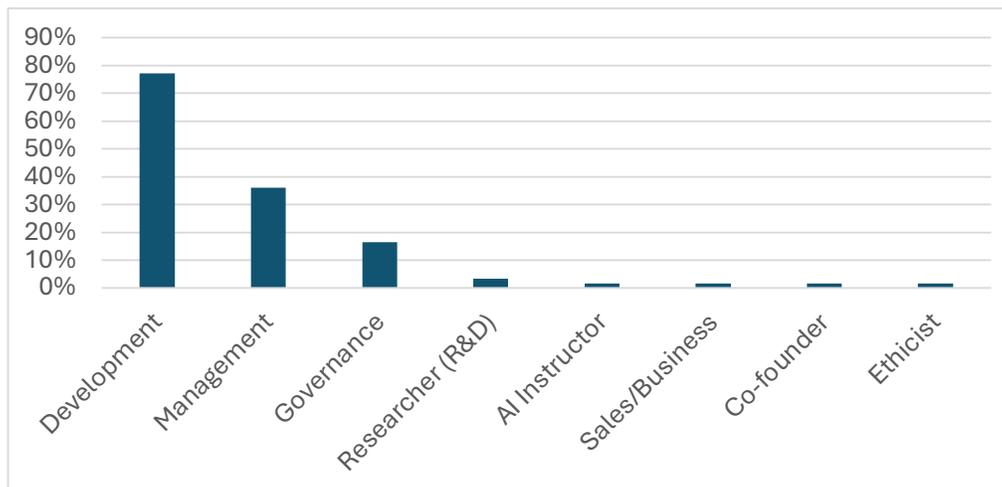

(c) Roles

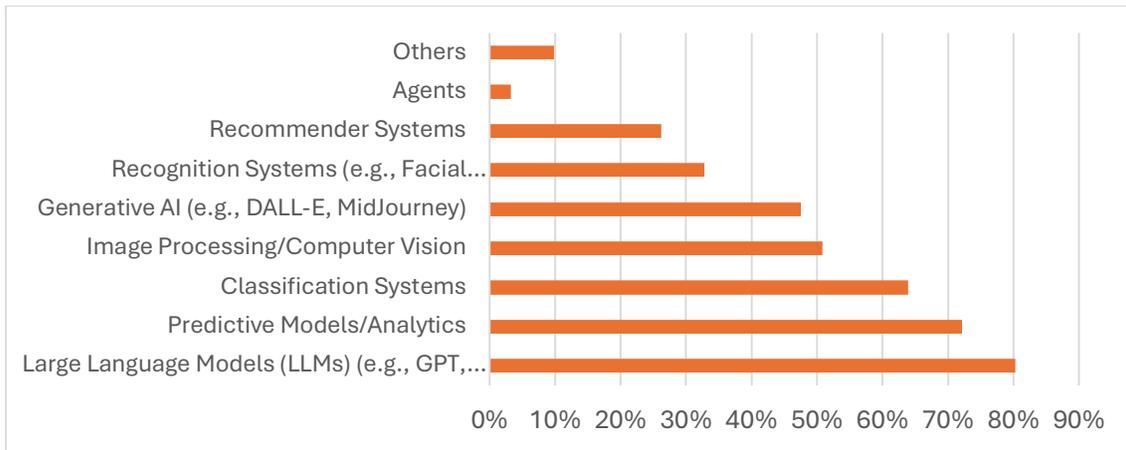

(d) Systems

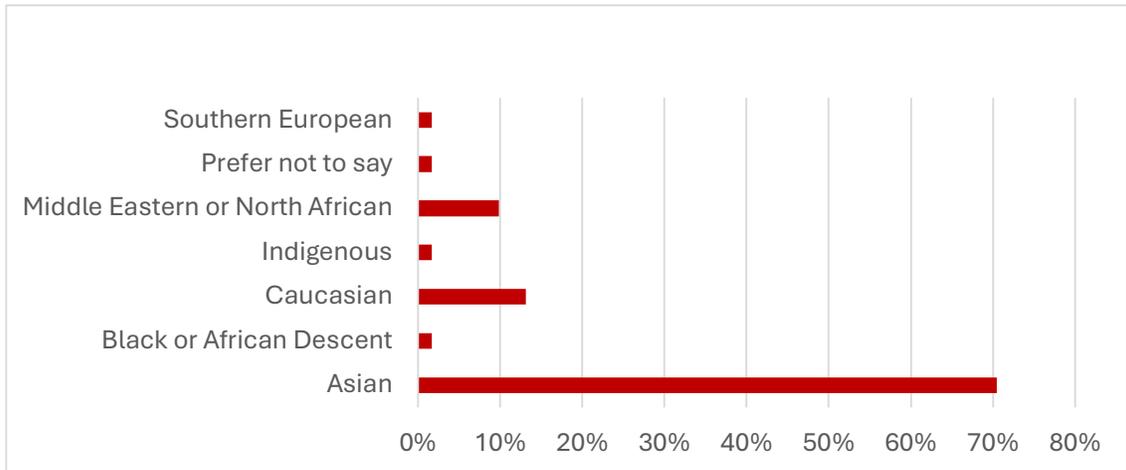

(e) Ethnicity

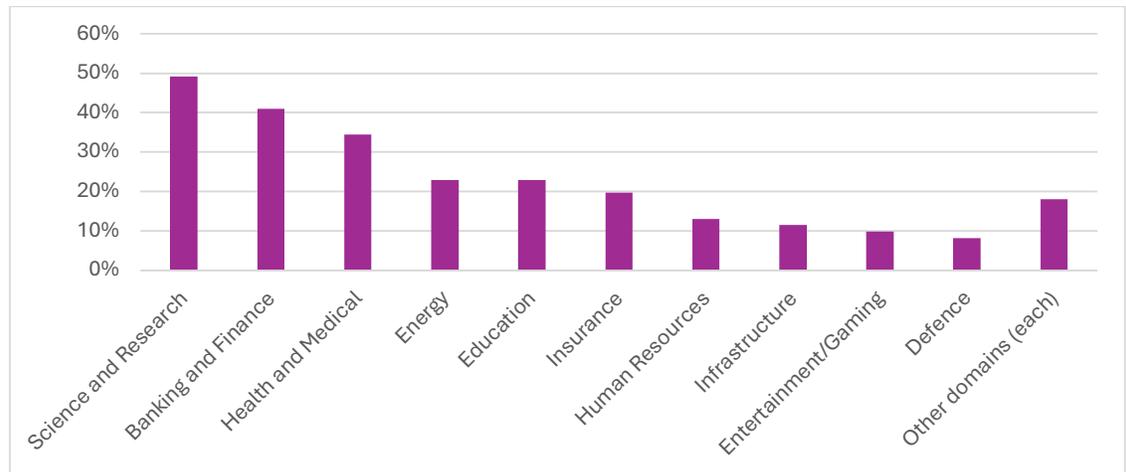

(f) Domains

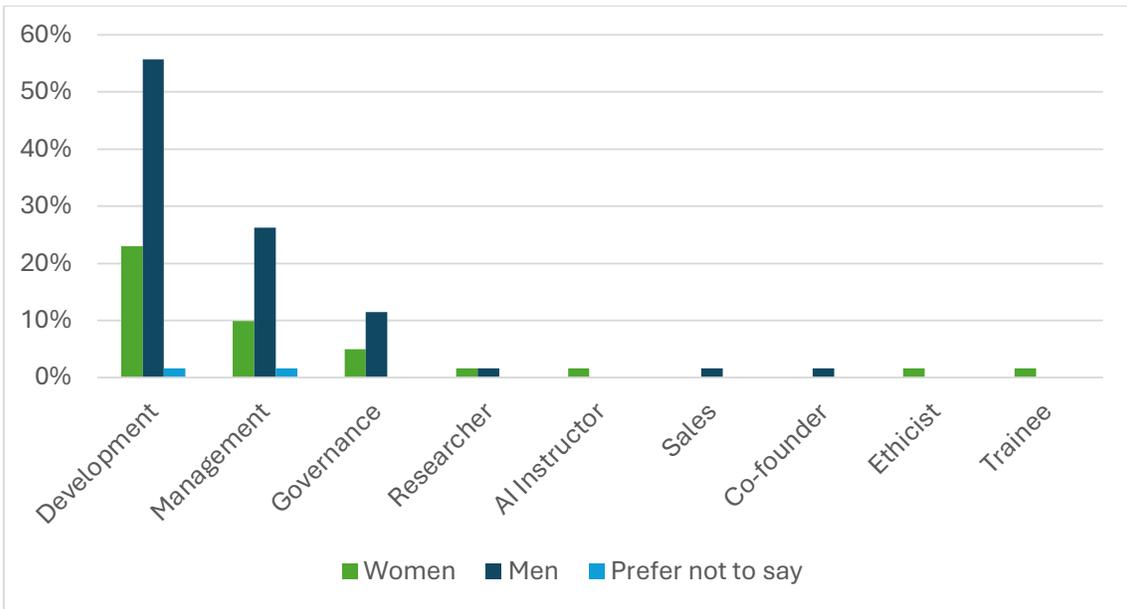
Fig. 11: Gender vs. Roles

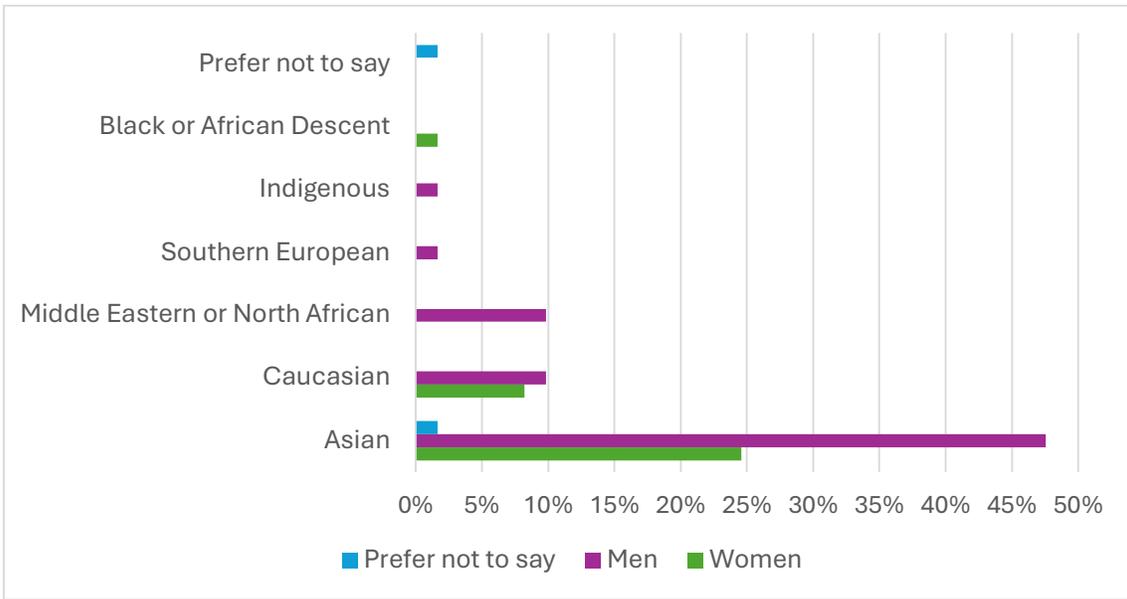
Fig. 12: Gender vs. Ethnicity

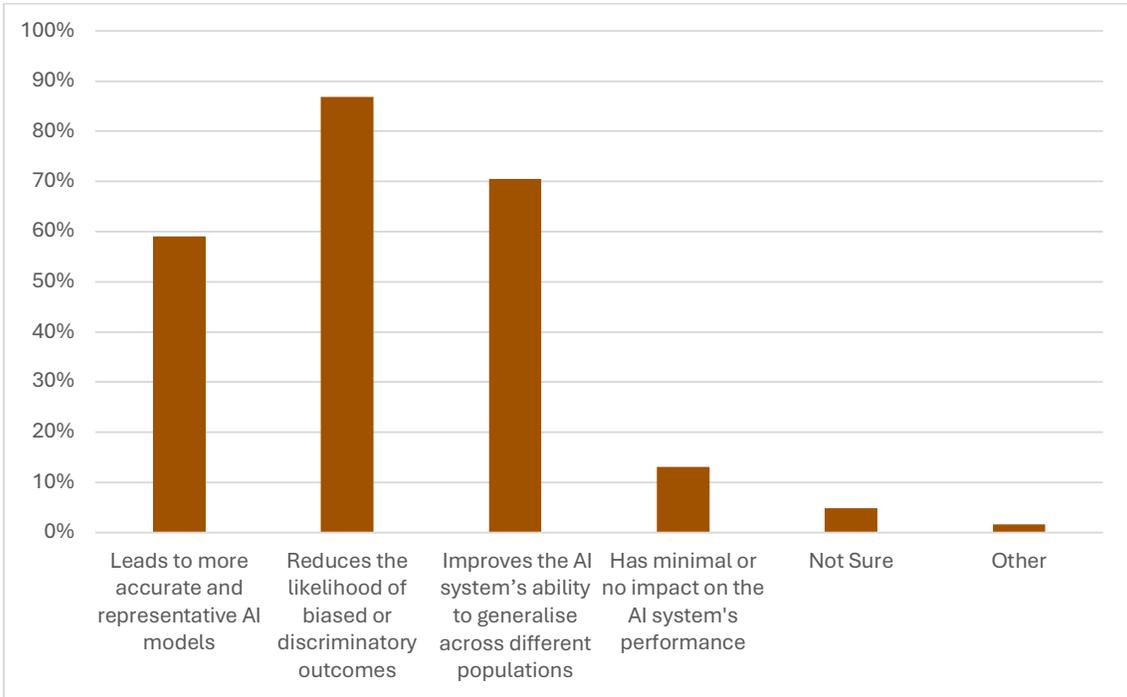

Fig. 13: Perceived Impact of Diverse and Inclusive Data

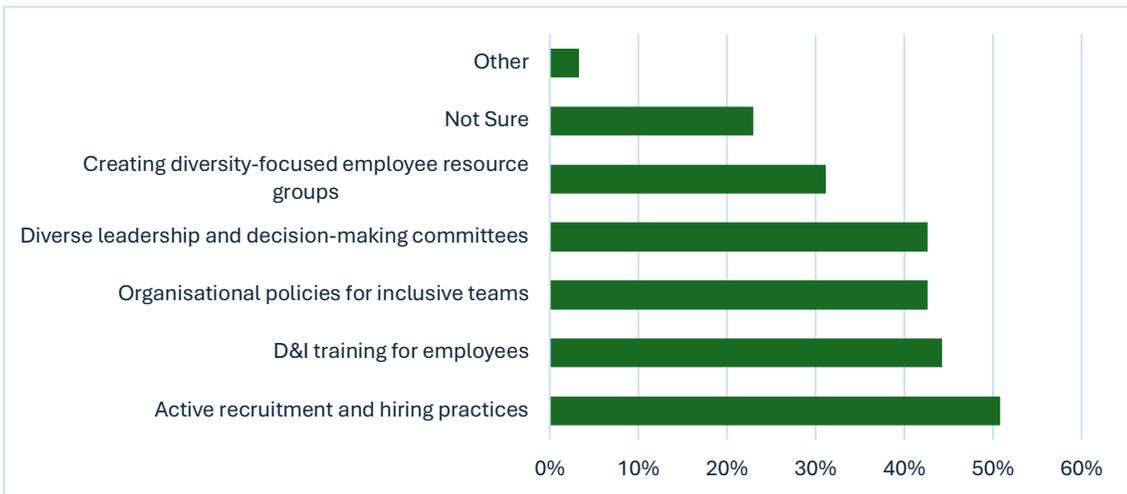

Fig. 14: Current Practices: Ensuring diverse and inclusive teams

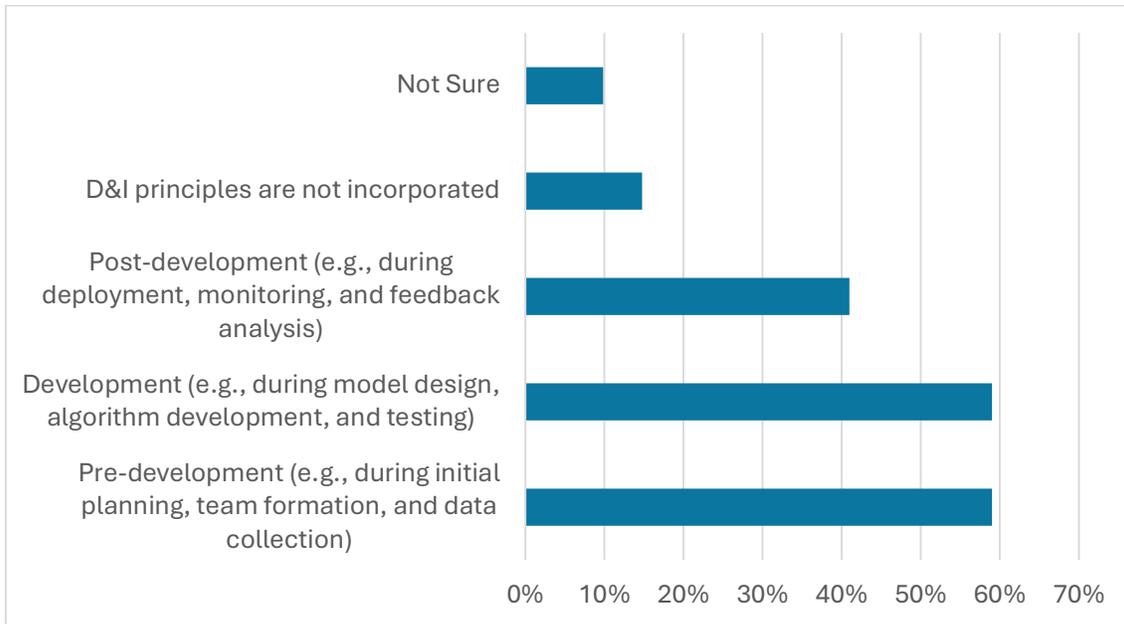

Fig. 15: Current Practices: D&I principles in AI lifecycle

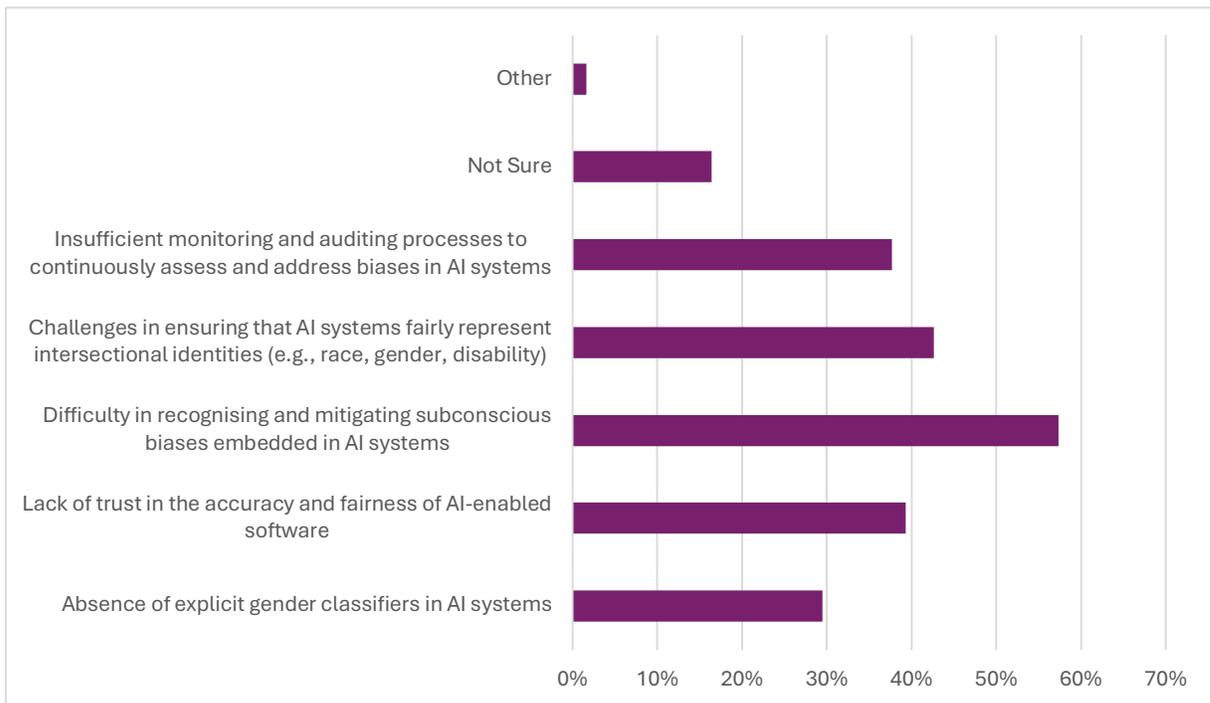

Fig. 16: Challenges in creating diverse and inclusive AI systems



\end{document}